\documentclass[12pt]{article}
\begin{document}
\newcommand{\beq}{\begin{equation}}
\newcommand{\eeq}{\end{equation}}
\newcommand{\beqa}{\begin{eqnarray}}
\newcommand{\eeqa}{\end{eqnarray}}
\newcommand{\beqar}{\begin{eqnarray*}}
\newcommand{\eeqar}{\end{eqnarray*}}
\newcommand{\al}{\alpha}
\newcommand{\be}{\beta}
\newcommand{\del}{\delta}
\newcommand{\D}{\Delta}
\newcommand{\eps}{\epsilon}
\newcommand{\ga}{\gamma}
\newcommand{\Ga}{\Gamma}
\newcommand{\ka}{\kappa}
\newcommand{\nn}{\nonumber}
\newcommand{\inn}{\!\cdot\!}
\newcommand{\h}{\eta}
\newcommand{\ii}{\iota}
\newcommand{\kk}{\varphi}
\newcommand\F{{}_3F_2}
\newcommand{\la}{\lambda}
\newcommand{\La}{\Lambda}
\newcommand{\na}{\prt}
\newcommand{\Om}{\Omega}
\newcommand{\om}{\omega}
\newcommand\dS{\dot{\cal S}}
\newcommand\dB{\dot{B}}
\newcommand\dG{\dot{G}}
\newcommand\ddG{\ddot{G}}
\newcommand\ddB{\ddot{B}}
\newcommand\ddP{\ddot{\phi}}
\newcommand\dP{\dot{\phi}}
\newcommand{\p}{\phi}
\newcommand{\sig}{\sigma}
\renewcommand{\t}{\theta}
\newcommand{\z}{\zeta}
\newcommand{\ssc}{\scriptscriptstyle}
\newcommand{\eg}{{\it e.g.,}\ }
\newcommand{\ie}{{\it i.e.,}\ }
\newcommand{\labell}[1]{\label{#1}} 
\newcommand{\reef}[1]{(\ref{#1})}
\newcommand\prt{\partial}
\newcommand\veps{\varepsilon}
\newcommand{\pol}{\varepsilon}
\newcommand\vp{\varphi}
\newcommand\ls{\ell_s}
\newcommand\cF{{\cal F}}
\newcommand\cA{{\cal A}}
\newcommand\cS{{\cal S}}
\newcommand\cT{{\cal T}}
\newcommand\cV{{\cal V}}
\newcommand\cL{{\cal L}}
\newcommand\cM{{\cal M}}
\newcommand\cN{{\cal N}}
\newcommand\cG{{\cal G}}
\newcommand\cH{{\cal H}}
\newcommand\cI{{\cal I}}
\newcommand\cJ{{\cal J}}
\newcommand\cl{{\iota}}
\newcommand\cP{{\cal P}}
\newcommand\cQ{{\cal Q}}
\newcommand\cg{{\it g}}
\newcommand\cR{{\cal R}}
\newcommand\cB{{\cal B}}
\newcommand\cO{{\cal O}}
\newcommand\tcO{{\tilde {{\cal O}}}}
\newcommand\bg{\bar{g}}
\newcommand\bb{\bar{b}}
\newcommand\bH{\bar{H}}
\newcommand\bX{\bar{X}}
\newcommand\bK{\bar{K}}
\newcommand\bA{\bar{A}}
\newcommand\bZ{\bar{Z}}
\newcommand\bxi{\bar{\xi}}
\newcommand\bphi{\bar{\phi}}
\newcommand\bpsi{\bar{\psi}}
\newcommand\bprt{\bar{\prt}}
\newcommand\bet{\bar{\eta}}
\newcommand\btau{\bar{\tau}}
\newcommand\bnabla{\bar{\nabla}}
\newcommand\hF{\hat{F}}
\newcommand\hA{\hat{A}}
\newcommand\hT{\hat{T}}
\newcommand\htau{\hat{\tau}}
\newcommand\hD{\hat{D}}
\newcommand\hf{\hat{f}}
\newcommand\hg{\hat{g}}
\newcommand\hp{\hat{\phi}}
\newcommand\hi{\hat{i}}
\newcommand\ha{\hat{a}}
\newcommand\hb{\hat{b}}
\newcommand\hQ{\hat{Q}}
\newcommand\hP{\hat{\Phi}}
\newcommand\hS{\hat{S}}
\newcommand\hX{\hat{X}}
\newcommand\tL{\tilde{\cal L}}
\newcommand\hL{\hat{\cal L}}
\newcommand\tG{{\widetilde G}}
\newcommand\tg{{\widetilde g}}
\newcommand\tphi{{\widetilde \phi}}
\newcommand\tPhi{{\widetilde \Phi}}
\newcommand\td{{\tilde d}}
\newcommand\tk{{\tilde k}}
\newcommand\tf{{\tilde f}}
\newcommand\ta{{\tilde a}}
\newcommand\tb{{\tilde b}}
\newcommand\tc{{\tilde c}}
\newcommand\tR{{\tilde R}}
\newcommand\teta{{\tilde \eta}}
\newcommand\tF{{\widetilde F}}
\newcommand\tK{{\widetilde K}}
\newcommand\tE{{\widetilde E}}
\newcommand\tpsi{{\tilde \psi}}
\newcommand\tX{{\widetilde X}}
\newcommand\tD{{\widetilde D}}
\newcommand\tO{{\widetilde O}}
\newcommand\tS{{\tilde S}}
\newcommand\tB{{\widetilde B}}
\newcommand\tA{{\widetilde A}}
\newcommand\tT{{\widetilde T}}
\newcommand\tC{{\widetilde C}}
\newcommand\tV{{\widetilde V}}
\newcommand\thF{{\widetilde {\hat {F}}}}
\newcommand\Tr{{\rm Tr}}
\newcommand\tr{{\rm tr}}
\newcommand\STr{{\rm STr}}
\newcommand\hR{\hat{R}}
\newcommand\M[2]{M^{#1}{}_{#2}}

\newcommand\bS{\textbf{ S}}
\newcommand\bI{\textbf{ I}}
\newcommand\bJ{\textbf{ J}}

\begin{titlepage}
\begin{center}

\vskip 2 cm
{\LARGE \bf Higher-derivative field redefinitions   \\ \vskip 0.75  cm  in the presence of boundary  }\\
\vskip 1.25 cm
   Mohammad R. Garousi\footnote{garousi@um.ac.ir}

\vskip 1 cm
{{\it Department of Physics, Faculty of Science, Ferdowsi University of Mashhad\\}{\it P.O. Box 1436, Mashhad, Iran}\\}
\vskip .1 cm
 \end{center}

\begin{abstract}
Recently it has been proposed   that the consistency with T-duality requires the effective action of string theory at order $\alpha'^n$ to satisfy the least action principle provided that the values of the massless fields and their  derivatives up to order $n$ are known on the boundary.  In this paper we speculate that  this boundary condition constrains the  field redefinitions and the corrections to the T-duality transformations in the presence of boundary, \eg at order $\alpha'$, the metric does not change,  and all other massless fields should change to include only the first derivative of the massless fields.

Using the above  restricted field redefinitions, we write all  gauge invariant bulk and boundary couplings in the bosonic string theory at order $\alpha'$ in a minimal scheme. Then using the assumption that the  effective action of string theory at the critical dimension is background independent, we fix the coefficients of the tree-level gauge invariant couplings  by imposing $O(1,1)$ symmetry when the background has a circle and  by imposing $O(d,d)$ symmetry when the background has $T^d$. These constraints fix the bulk action up to an overall factor, and the boundary action up to two parameters.  By requiring the gravity couplings in the boundary action to be  consistent with those in the Chern-Simons gravity, the two boundary parameters are also fixed. Up to a restricted  field redefinition, the bulk and boundary couplings are exactly those in the  K.A. Meissner action and its  corresponding boundary action.
\end{abstract}
\end{titlepage}

\section{Introduction}

string theory is expected  to be a background independent theory of quantum gravity. The spectrum of the free string at the critical dimension which is  26 for the bosonic string theory and 10 for the superstring and heterotic string theories,  includes  a finite number of massless and a tower of infinite number of  massive  excitations. In the interacting theory and at   low energies, however,  the massive modes are integrated out to produce an effective action which includes only the massless fields.  The effective action of string theory on an arbitrary  background  has a double expansions: the genus-expansion and the stringy-expansion. The   genus-expansion  includes  the  classical sphere-level  and a tower of  quantum loop-level  corrections. The  stringy-expansion  is an expansion in terms of  higher derivative couplings at each loop-level.  The classical    effective action of string theory  on an arbitrary open manifold  has both bulk and boundary couplings,  \ie $\bS_{\rm eff}+\prt\!\! \bS_{\rm eff}$. At the critical dimension, it has  the following  higher-derivative or  $\alpha'$-expansion:
\beqa
\bS_{\rm eff}&=&\sum^\infty_{m=0}\alpha'^m\bS_m=\bS_0+\alpha' \bS_1 +\alpha'^2 \bS_2+\alpha'^3 \bS_3+\cdots \labell{seff}\\
\prt\!\!\bS_{\rm eff}&=&\sum^\infty_{m=0}\alpha'^m\prt\!\!\bS_m=\prt\!\!\bS_0+\alpha' \prt\!\!\bS_1+\alpha'^2 \prt\!\!\bS_2+\alpha'^3 \prt\!\!\bS_3 +\cdots \nn
\eeqa
The leading order bulk action $\!\!\bS_0$   includes the Hilbert-Einstein term in the critical dimension and   the boundary action $\prt\!\!\bS_0$  includes the corresponding Hawking-Gibbons term \cite{York:1972sj,Gibbons:1976ue}. These actions and their appropriate  higher derivative extensions may be  found by requiring the effective actions to be invariant under the gauge symmetries corresponding to the massless fields and by imposing various constraints from  the global symmetries/dualities of string theory.

The Einstein action  is background independent in the sense that only the gauge symmetry corresponding to the metric is required to specify the theory. 
We expect that string theory  effective action at the critical dimension which is a higher-derivative extension of the Einstein term, to be background independent too. Unlike the Einstein action which has only one coupling, however, there are many gauge invariant couplings in the effective action of string theory at each order of $\alpha'$. The background independence  requires the coefficients of all  gauge invariant couplings  to be independent of the geometry of the spacetime. At the higher orders of $\alpha'$, there is also the complication   that the effective action has the freedom of the higher-derivative field redefinitions \cite{Metsaev:1987zx}. If one could fix the parameters, up to field redefinitions, for a specific compact geometry in which the lower-dimensional   action enjoys a specific global symmetry, then those parameters would be  valid for any other geometry in which the effective action may have  no compact sub-manifold and no symmetry. 

One of the most exciting discoveries in the  perturbative  string theory is    T-duality \cite{Giveon:1994fu,Alvarez:1994dn}  which appears  when one compactifies the theory  on a torus, \eg the spectrum of the free string  on torus $T^d$ is invariant under   $O(d,d,Z)$ transformations. In the interacting theory and after integrating out the massive modes,   T-duality  should appear as symmetry in the dimensional reduction of the  effective actions. For the closed spacetime manifolds, it has been shown in \cite{Sen:1991zi,Hohm:2014sxa} that the dimensional reduction of the classical effective actions of the bosonic and heterotic string theories on a torus $T^d$ are in fact invariant under  $O(d,d,R)$ transformations. We expect there should be such symmetry  even when the string lives in an open spacetime manifold which has boundary. Using the background independent assumption, then one may consider the background which has the torus $T^n$ for  $1\leq n\leq d$, to study the allowed  gauge invariant couplings in the effective action at the critical dimension. That is, the requirement  that the dimensional reduction of the classical effective action to have the symmetry $O(n,n,R)$, strongly constrains   the parameters of the gauge invariant couplings  in both bulk and boundary  actions.

For the closed spacetime manifolds, if one  considers a  background which includes   only one circle $S^{(1)}$, then the requirement  that the dimensional reduction of the classical effective action of the bosonic string theory to have  the $O(1,1)$ symmetry,  fixes  all  couplings in the critical dimension    up to, at most, one  parameter at  each order of $\alpha'$.  In fact,  the constraints from the $Z_2$-subgroup of $O(1,1)$ have been used  in \cite{Garousi:2019wgz,Garousi:2019mca} to fix the classical   effective actions of  the bosonic string theory at  orders $\alpha'\,, \alpha'^2$ up to one parameter, \ie couplings at orders $\alpha'$ and $\alpha'^2$ have the overall factors $a_1$ and $a_1^2$, respectively. The $Z_2$-constraints also fix the NS-NS couplings of the  type II superstring theories at order $\alpha'^3$ up to another parameter  $a_2\sim\z(3)$ \cite{Razaghian:2018svg,Garousi:2020mqn,Garousi:2020gio,Garousi:2020lof}. The background independence then indicates that  the couplings found in this way are valid for any other geometry.  In fact, it has been observed in \cite{Garousi:2020gio,Garousi:2020lof} that the NS-NS couplings are  fully consistent with the sphere-level S-matrix element of four NS-NS vertex operators  in flat spacetime. Moreover, it has been shown in \cite{Garousi:2021ikb,Garousi:2021ocs} that the  above couplings are  fully consistent with the $O(d,d)$ symmetry when the background has torus $T^d$.

The reason that the $O(1,1)$ symmetry can not  fully fix all couplings in the classical effective action of the bosonic string theory  
up to one overall factor  
 is that the full T-duality transformations are the Buscher rules \cite{ Buscher:1987sk,Buscher:1987qj} plus  derivative corrections at all orders of $\alpha'$ \cite{Kaloper:1997ux}. Even though the $O(1,1)$ symmetry can fix all couplings at orders $\alpha',\alpha'^2$ up to one parameter, however, it can not fix the couplings at order  $\alpha'^3$, up to the same parameter. 
In fact, if one extends the calculations in \cite{Garousi:2019mca} to the order $\alpha'^3$, one would find that the couplings at orders $\alpha'^0$,  $\alpha'$ and $\alpha'^2$ are related to  some of the couplings at order $\alpha'^3$ by the T-duality transformations at orders $\alpha'^3$, $\alpha'^2$ and $\alpha'$, respectively. These couplings  carry the same parameter $a_1$  which appears in the couplings at orders $\alpha',\alpha'^2$. In other words, they all belong to one   T-dual multiplet. However, there are couplings at this order that are not connected to the couplings at order $\alpha',\alpha'^2$ by the T-duality transformations. They are connected only to the couplings at order $\alpha'^0$ by the T-duality transformations at order $\alpha'^3$. These couplings  belong to another T-dual multiplet with a different parameter $a_2$.

In other words, if one tries to find  the classical effective action of  the bosonic string theory at order $\alpha'^3$ by the S-matrix method, instead of the T-duality method, one would find it has two factors $a_1,a_2$, \ie $\!\bS_3=a_1^3\!\bS_3^1+a_2\!\bS_3^2$. One factor which is resulted from the expansion of the tachyonic pole in the amplitude,  should be the same as the one appearing in the couplings at order $\alpha',\alpha'^2$, and another one which is  resulted from the expansion of the massive poles in the amplitude, is proportional to  $\z(3)$ \cite{Jack:1989vp}. Hence, there are two T-dual multiplets in the bosonic string theory at order $\alpha'^3$: one with coefficient $a_1^3$ and another one with coefficient $a_2\sim \z(3)$. The multiplet with coefficient $\z(3)$ appears  also in the type II supersting theory.   At  order $\alpha'^4$, there are non-zero couplings in the bosonic string theory. However,  there are no  couplings at this order  in the superstring theory \cite{Stieberger:2009rr}. Hence, there are still two T-dual multiplets at this order: one with coefficient $a_1^4$ and another one with coefficient $a_1\z(3)$. There is no new parameter at order $\alpha'^4$ in this case.  At order $\alpha'^5$, there are couplings $R^6$ in the superstring theory with coefficient $\z(5)$ \cite{Stieberger:2009rr}, hence, there  are  three T-dual multiplets at order $\alpha'^5$ in the bosonic string: one with coefficient $a_1^5$, one with coefficient $a_1^2\z(3)$ and another one with coefficient $\z(5)$. This means that the $O(1,1)$ symmetry can fix the couplings at order $\alpha'^5$ up to three parameters. Two of them already appeared at the lower orders of $\alpha'$ and  one new parameter appears at order $\alpha'^5$.

In general, there is no coupling in the superstring theory with coefficient $\z(2k)$ \cite{Stieberger:2009rr} which can be seen by studying the $\alpha'$-expansion of the sphere-level four point functions. As a result, the  T-dual multiplets in the effective action of the bosonic string theory should have an expansion in terms of powers of $(a_1^{m_1}\z(2k+1)^{m_2}\z(2k+3)^{m_3}\cdots)\alpha'^{m_1+m_2(2k+1)+m_3(2k+3)+\cdots}$ where $k=1,2,\cdots$ and $m_1,m_2,\cdots=0,1,2,\dots$.   For $m_2=m_3=\cdots=0$ there is one T-dual multiplet with parameter $a_1$. For $m_1=m_3=\cdots=0$ and $k=1$, there is one T-dual multiplet with parameter  proportional to $\z(3)$. Similarly for other values of $k,m_i$.  Schematically, the bulk action \reef{seff} has the following expansion in terms of the  T-dual multiplets:
\beqa
\bS_{\rm eff}&=&\sum^\infty_{n=0} T_n=T_0+ T_1+T_2+T_3+T_4+T_5\cdots \labell{Teff}
\eeqa
  The T-dual multiplets in the bosonic string theory on a closed manifold have the following structures:
 \beqa
 T_0&=&\bS_0\nn\\
 T_1&=&a_1\alpha' \bS_1+a_1^2\alpha'^2\bS_2+a_1^3\alpha'^3\bS_3^1+a_1^4\alpha'^4\bS_4^1+a_1^5\alpha'^5\bS_5^1+\cdots\nn\\
  T_2&=&\z(3)\alpha'^3 \bS_3^2+\z(3)^2\alpha'^6\bS_6^2+\z(3)^3\alpha'^{9}\bS_{9}^2+\z(3)^4\alpha'^{12}\bS_{12}^2+\cdots\nn\\
  T_3&=&a_1\z(3)\alpha'^4\bS_4^3+a_1^2\z(3)^2\alpha'^8\bS_8^3+a_1^3\z(3)^3\alpha'^{12}\bS_{12}^3+a_1^4\z(3)^4\alpha'^{16}\bS_{16}^3+\cdots\nn\\
   T_4&=&a_1^2\z(3)\alpha'^5\bS_5^4+a_1^4\z(3)^2\alpha'^{10}\bS_{10}^4+a_1^6\z(3)^3\alpha'^{15}\bS_{15}^4+a_1^8\z(3)^4\alpha'^{20}\bS_{20}^4+\cdots\nn\\
    T_5&=&\z(5)\alpha'^5\bS_5^5+\z(5)^2\alpha'^{10}\bS_{10}^5+\z(5)^3\alpha'^{15}\bS_{15}^5+\z(5)^4\alpha'^{20}\bS_{20}^5+\cdots\labell{bulkT}\\
  \vdots\nn
 \eeqa
 where $a_1\sim 1$.  In the type II superstring theory, $a_1=0$ and there are T-dual multiplets which include the R-R couplings.
  If the background has no boundary in which the total derivative terms  can be ignored, then each of the above multiplets should be  invariant under the full T-duality transformations. We expect that the $O(1,1)$-constraint  to fix all couplings in the above  T-dual multiplets, up to the  overall factors $a_1,\z(3),a_1\z(3),a_1^2\z(3),\z(5),\cdots$. The couplings in the multiplet $T_1$ at orders $\alpha'$ and $\alpha'^2$ in a particular minimal scheme have been found in \cite{Garousi:2019wgz,Garousi:2019mca}.   The couplings in the multiplet $T_2$ at order $\alpha'^3$ in a particular minimal scheme have been also found in \cite{Garousi:2020gio,Garousi:2020lof}. At the present time, the higher derivative terms in the full  T-duality transformations are not known. They depend on the scheme of the gauge invariant couplings \cite{Garousi:2019wgz}. They can be  found at each order of $\alpha'$, up to overall factors,  by imposing the effective action at that order to be invariant under the  $O(1,1)$ symmetry. Hence, the T-duality constraint  can not predict the overall factors $a_1,\z(3),a_1\z(3),a_1^2\z(3),\z(5),\cdots$ which appear in the full T-duality transformations and in  the above T-dual multiplets. Since each multiplet includes four-field couplings, however,   these factors  may be found  from the $\alpha'$-expansion of the sphere-level S-matrix element of four graviton vertex operators.

When the background  has boundary, one should keep the total derivative terms before and after reduction and use  the Stokes's theorem to transfer them  to the boundary. They dictate that the invariance under the T-duality transformations requires some couplings on the boundary  as well  \cite{Garousi:2019xlf}.   Hence, the bulk T-dual multiplets \reef{bulkT} should be accompanied  with appropriate boundary couplings to be fully invariant under the T-duality. Schematically, the boundary action \reef{seff} should have  the following expansion in terms of the boundary T-dual multiplets:
\beqa
\prt\!\!\bS_{\rm eff}&=&\sum^\infty_{n=0} \prt T_n=\prt T_0+ \prt T_1+ \prt T_2+\prt T_3+ \prt T_4+ \prt T_5+\cdots \labell{Teffb}
\eeqa
   The boundary  mutiplets corresponding to the bulk multiplets \reef{bulkT} have the following structures:
    \beqa
\prt T_0&=&\prt\!\!\bS_0\nn\\
\prt T_1&=&a_1\alpha' \prt\!\!\bS_1+a_1^2\alpha'^2\prt\!\!\bS_2+a_1^3\alpha'^3\prt\!\!\bS_3^1+a_1^4\alpha'^4\prt\!\!\bS_4^1+a_1^5\alpha'^5\prt\!\!\bS_5^1+\cdots\nn\\
\prt T_2&=&\z(3)\alpha'^3\prt\!\! \bS_3^2+\z(3)^2\alpha'^6\prt\!\!\bS_6^2+\z(3)^3\alpha'^{9}\prt\!\!\bS_{9}^2+\z(3)^4\alpha'^{12}\prt\!\!\bS_{12}^2+\cdots\nn\\
\prt T_3&=&a_1\z(3)\alpha'^4\prt\!\!\bS_4^3+a_1^2\z(3)^2\alpha'^8\prt\!\!\bS_8^3+a_1^3\z(3)^3\alpha'^{12}\prt\!\!\bS_{12}^3+a_1^4\z(3)^4\alpha'^{16}\prt\!\!\bS_{16}^3+\cdots\nn\\
\prt T_4&=&a_1^2\z(3)\alpha'^5\prt\!\!\bS_5^4+a_1^4\z(3)^2\alpha'^{10}\prt\!\!\bS_{10}^4+a_1^6\z(3)^3\alpha'^{15}\prt\!\!\bS_{15}^4+a_1^8\z(3)^4\alpha'^{20}\prt\!\!\bS_{20}^4+\cdots\nn\\
\prt  T_5&=&\z(5)\alpha'^5\prt\!\!\bS_5^5+\z(5)^2\alpha'^{10}\prt\!\!\bS_{10}^5+\z(5)^3\alpha'^{15}\prt\!\!\bS_{15}^5+\z(5)^4\alpha'^{20}\prt\!\!\bS_{20}^5+\cdots\labell{boundT}\\
  \vdots\nn
 \eeqa
The combination of the bulk and the boundary multiplets, \ie $ T_i+\prt T_i$, are then invariant under the T-duality transformations. In other words,  neither the bulk multiplets nor the boundary multiplets are invariant separately under the T-duality transformations. Their anomalies cancel each other in the T-dual multiplets $ T_i+\prt T_i$.  This constrains the  parameters in the  general gauge invariant couplings in the boundary action \reef{boundT}. However, there are   boundary couplings in \reef{boundT} which are invariant under the T-duality transformations with no anomaly. They form T-dual boundary multiplets which have no bulk partner. To constrain these multiplets, one may use the background which has torus $T^d$ for which  the cosmological  reduction of the effective action should have the  $O(d,d)$ symmetry \cite{Sen:1991zi,Hohm:2014sxa}.

 To impose the $O(1,1)$-constraint on the gauge invariant couplings in  the closed spacetime manifold, one should first use the most general field redefinitions  to find the independent couplings in a minimal scheme \cite{Metsaev:1987zx} in which the number of independent gauge invariant couplings is minimum, and then should use the most general corrections to the Buscher rules \cite{ Buscher:1987sk,Buscher:1987qj} to fix  the T-dual multiplets in \reef{bulkT} at order $\alpha'^n$ up to the overall factors at each order of $\alpha'$.   It turns out that in the presence of boundary, however, if one uses the most general field redefinitions and the most general corrections to the T-duality transformations,  then the couplings in the minimal scheme are not invariant  under  the T-duality transformations! It may indicate that in the present of boundary, one is not allowed to use the most general  field redefinitions and the most general corrections to the Buscher rules.

It has been observed in  \cite{Myers:1987yn} that only the gravity couplings in the  Euler characters satisfy the least action principle with the usual boundary condition that  the values of the metric are known on the boundary. On the other hand, it has been shown in \cite{Garousi:2021cfc} that the Euler character at order $\alpha'$  is not consistent with $O(1,1)$ nor with $O(d,d)$ symmetries. Hence, the T-duality dictates that in the least action principle, not only the values of the massless fields but also the values of some derivatives of the massless field must be known on the boundary.  In fact, as it has been argued in  \cite{Myers:1987yn}, the least action principle in the bosonic string field theory produces the correct string field equations of motion with the usual  boundary condition that the values of the string field are known on the boundary. The string field includes the massless fields and infinite tower of the massive fields. Hence, in the least action the values of the massless fields an all massive fields are known on the boundary. When the massive fields are  integrated out to produce the effective action, the values of the massive fields on the boundary should appear in the effective action as the values of the derivatives of the massless fields on the boundary. It has been proposed in \cite{Garousi:2021cfc} that  to extremize the effective action at order $\alpha'^n$ for $n>0$,   not only the massless fields $\Psi$ but also their derivatives  up to order $n$, \ie $\nabla\Psi,\cdots, \nabla^n\Psi$ must be  known on the boundary. It means that in the least action principle $\delta(\!\!\bS_{eff}+\prt\!\!\bS_{eff})=0$, the variation of the massless fields and the variation of the derivatives of the massless fields up to order $n$ must be  zero on the boundary, \ie $\delta\Psi=\delta\nabla\Psi=\cdots=\delta\nabla^n\Psi=0$  on the boundary. However, the variation of the derivatives of the massless fields at the higher orders  are not zero on the boundary, \ie $\delta\nabla^{n+1}\Psi\neq 0,\delta\nabla^{n+2}\Psi\neq 0$, and so on.  These boundary values  do not constrain the couplings in the bulk effective action. They, however,  constrain the couplings in  the boundary actions.  In fact, using  the Stokes's theorem to remove the total derivative terms in the bulk action to the boundary and using the appropriate Bianchi identities, one can show that the  gauge invariant  bulk couplings at order $\alpha'^n$ have no derivative terms   $\nabla^{n+2}\Psi$ and higher. Then in extremizing the bulk action, using the Stokes's theorem, one finds  no term with variation   $\delta\nabla^{n+1}\Psi$ and higher. Hence, the extremization  $\delta(\!\!\bS_{eff})=0$ produces equations of motion with no constraint on the parameters in the bulk effective action. However, in extremizing the boundary action $\delta(\prt\!\!\bS_{eff})=0$, the variations   $\delta\nabla^{n+1}\Psi$, $\delta\nabla^{n+2}\Psi$ and higher may  appear on the boundary which are not zero.   The extremization  $\delta(\prt\!\!\bS_{eff})=0$ then constrains the parameters in the boundary action \reef{boundT}.

 Therefore, when the background  has boundary, because of the above boundary conditions,  one is  allowed to use only a restricted field redefinitions in which the values of the massless fields and their derivatives to order $\alpha'^n$ on the boundary  remain invariant under the field redefinitions.  The field redefinitions should not change the known values to the unknown values. In other words,   the least action principle forces the variations of fields and their derivatives up to order $n$ to be zero on the boundary, \ie   $\delta\Psi=\delta\nabla\Psi=\cdots=\delta\nabla^n\Psi=0$   and   $\delta\nabla^{n+1}\Psi\neq 0,\delta\nabla^{n+2}\Psi\neq 0, \cdots$, on the boundary. The field redefinitions should not change  $\delta\Psi,\,\delta\nabla\Psi\,,\cdots\,, \delta\nabla^n\Psi$ which are zero,  to  $\delta\nabla^{n+1}\Psi\,,\delta\nabla^{n+2}\Psi\,, \cdots$ which are non-zero on the boundary.  For example, for the effective action at order $\alpha'$, the values of the massless fields and their first derivatives are known on the boundary and the values of their higher derivatives are not known. The  field redefinitions at order $\alpha'$ are applied on the effective action at order $\alpha'^0$ to produce couplings at order $\alpha'$.  At  order $\alpha'^0$, only the values of the massless fields are known on the boundary. Using the fact that the first derivative of dilaton and $B$-field appear in the leading order effective action, one realizes that in the least action, using the Stokes's theorem, only the variation of the dilaton and $B$-field appears  on the boundary. These variations are zero on the boundary at the leading order.  The field redefinition of these fields which includes only the first derivative of the massless fields, are then allowed because they produce the variations for the first derivative of the massless fields. Such variations  are allowed for the effective action at order $\alpha'$ because they are zero at this order. On the other hand, the second derivative of metric appears in the leading order effective action. In the variation of the action, the variation of the second derivative of the metric appears. Then upon the use of the Stokes's theorem, it produces the  variation of the first derivative of metric on the boundary. If one uses a field redefinition  for the metric which includes the first derivative of the massless fields, then it would produce variation of the second derivative of the massless field on the boundary which are not allowed because the variation of the second derivative of the massless fields are not zero on the boundary for the effective action at order $\alpha'$. Hence, the correct field redefinitions at order $\alpha'$ do  not  allow to change the metric and  allow to change the dilaton and $B$-field to include only the first derivative of the massless fields.  Similarly for the field redefinition for the effective actions at higher orders of $\alpha'$. Using the above restricted field redefinitions, we will find that there are 17 independent bulk couplings at order $\alpha'$.

If the background  has boundary and   a compact sub-manifold independent  of the boundary, then the dimensionally  reduced  action at order $\alpha'^n$ should satisfy the least action principle in the base space with the  boundary conditions that the values of the base space massless fields and their derivatives up to order $n$ are known on the boundary. Hence, the higher-derivative field redefinitions in the base space  should be restricted as those specified in the previous paragraph, \ie they should not change the  data on the boundary. If the compact manifold is a circle, then  the T-duality transformations in the presence of boundary should be the Buscher rules \cite{ Buscher:1987sk,Buscher:1987qj} plus some restricted  higher derivative corrections, \eg at order $\alpha'$, the base space metric receives no corrections, and all other massless fields, except the torsion,  should receive corrections which include only the first derivative of the massless fields. The torsion 3-form in the base space however is not the exterior  derivative of a 2-form. Hence its variation is not the derivative of the variation of a 2-form. As a result, the corrections to the torsion 3-form may involve the second derivative of the massless fields without changing the boundary conditions on the massless fields and their first derivative. Similarly for the corrections  at the higher orders of $\alpha'$. Hence, in applying the $O(1,1)$-constraint on the independent gauge invariant couplings in the presence of the boundary,  one should use the above restricted T-duality transformations.

When one uses the cosmological reduction on the classical bulk effective action \reef{seff}, the resulting one-dimensional effective action has $O(d,d)$ symmetry provided that  the  $O(d,d)$ transformations  receive higher derivative corrections \cite{Meissner:1996sa}.
Using the most general corrections for the $O(d,d)$-transformations including corrections for the lapse function,  and  using  the integration by parts, it has been shown in \cite{Hohm:2015doa,Hohm:2019jgu} that the cosmological reduction of the bulk action \reef{seff} at order $\alpha'$ and higher, can be written in a scheme in which only  the first time-derivative of the generalized metric ${\cal S}$ appears, \ie the $nonlocal$ cosmological action becomes $local$ in the new variables.  It has been shown in  \cite{Garousi:2020gio,Garousi:2020lof} that the cosmological reduction of the couplings at order $\alpha'^2,\alpha'^3$ that have been found by the $O(1,1)$-constraint, can be written in terms of only traces of $\dot{{\cal S}}$. 
In the presence of boundary, however, the lapse function is the unit vector orthogonal  to the boundary. Hence, one should not use corrections  for the laps function any more. Moreover, in the presence of the boundary the higher derivative corrections to the $O(d,d)$-transformations should be restricted to those which do not change the data on the boundary, \eg the corrections at order $\alpha'$ should include only the first time-derivative of the massless fields.

In studying the $O(d,d)$ symmetry of  the cosmological reduction of the boundary action in \reef{seff}, one has to take into account the one-dimensional total derivative terms  that are needed to write the cosmological bulk couplings in $O(d,d)$-invariant form.  The total derivative terms, in general,  are not invariant under the $O(d,d)$ transformations, hence, they put some constraints on the parameters in the cosmological reduction of the boundary couplings. In other words, the cosmological reduction of the boundary couplings are anomalous under the $O(d,d)$-transformations such that their anomalies are canceled with the anomalies of the total derivative terms. In this case also there are boundary couplings that their cosmological reduction are invariant under the $O(d,d)$ symmetry with no anomaly. It has been speculated in \cite{Garousi:2021cfc} that in the scheme that the $O(d,d)$-invariant couplings are in terms of the first derivative of the generalized metric and  dilaton, the cosmological boundary action should be  zero,  \ie
\beqa
\prt\!\!\bS_{k}^{c}&=&0\labell{zeroS}
\eeqa
The above discussion is valid not only for the cosmological reduction whose boundary is spacelike, \ie $n^2=-1$, but also for any one-dimensional reduction whose boundary is timelike, \ie $n^2=1$. At the leading order which has no field redefinitions freedom, the explicit calculation confirms the above constraint. The above equation  constrains the parameters in the gauge invariant couplings in the boundary action $\prt\!\!\bS_k$ in \reef{boundT}. 

The remainder  of the paper is as follows:  In section 2, using the restricted field redefinitions, removing the total derivative terms from the bulk to the boundary and using the Bianchi identities,  we find that there are at least 17 independent gauge invariant couplings at order $\alpha'$ in the bulk. We write them in a specific minimal scheme which includes 17 bulk couplings, and write  the 38 independent  gauge invariant boundary couplings at order $\alpha'$ which have been found in \cite{Garousi:2021cfc}.  In section 3, using the background independence assumption, we consider the background which has a boundary and one circle, and use the dimensional reduction to find the corresponding couplings in the base space. We then  impose the $O(1,1)$ symmetry with the above restricted T-duality transformations, on the reduced actions to constrain the parameters in the actions.     In section 4, we consider the background which has a boundary and the torus $T^d$, and use the cosmological/one-dimensional  reduction to find the one-dimensional bulk action and the zero-dimensional boundary action. We then  impose the $O(d,d)$ symmetry with the above restricted corrections to the $O(d,d)$ transformations, on the resulting bulk action and the constraint \reef{zeroS} on the boundary action  to further constrain the remaining parameters.    The above constraints fix the bulk action up to an overall factor, and the boundary action up to two parameters. The gravity couplings in the bulk action are those in the Gauss-Bonnet gravity. By requiring  the gravity couplings on the boundary action to be  consistent with those in the Chern-Simons form, the two boundary parameters are also fixed. In section 6, we compare the bulk and boundary actions that we have found in this paper with the  Meissner action   and its  corresponding boundary action that has been recently found by the similar constraints. We show that they are related by a restricted field redefinitions. In section 7, we briefly discuss our results.

\section{Gauge invariance couplings  at order $\alpha'$}

The classical effective action of the bosonic string theory on an open manifold has both bulk and boundary actions. At the leading order these actions in the string frame are
\beqa
\bS_0+\prt\!\!\bS_0
=-\frac{2}{\kappa^2}\Bigg[ \int_M d^{26}x \sqrt{-G} e^{-2\Phi} \left(  R + 4\nabla_{\mu}\Phi \nabla^{\mu}\Phi-\frac{1}{12}H^2\right)+ 2\int_{\prt M} d^{25}\sigma\sqrt{| g|}  e^{-2\Phi}K\Bigg]\labell{baction}
\eeqa
where $G$ is determinant of the bulk metric $G_{\mu\nu}$ and boundary is specified by the functions $x^\mu=x^\mu(\sigma^{\tilde{\mu}})$. In the boundary term, $g$ is determinant of the induced metric on the boundary
\beqa
g_{\tilde{\mu}\tilde{\nu}}&=& \frac{\prt x^\mu}{\prt \sigma^{\tilde{\mu}}}\frac{\prt x^\nu}{\prt \sigma^{\tilde{\nu}}}G_{\mu\nu}\labell{indg}
\eeqa
and $K$ is the trace of the extrinsic curvature. The normal vector to the boundary is $n^\mu$. It is  outward-pointing (inward-pointing) if the boundary is spacelike (timelike).  

At order $\alpha'$ these actions in terms of their Lagrangians  are
\beqa
\bS_1=-\frac{2}{\kappa^2}\int_M d^{26} x\sqrt{-G} e^{-2\Phi}\mathcal{L}_1 ;\qquad\prt\!\!\bS_1=-\frac{2}{\kappa^2}\int_{\prt M} d^{25} \sigma\sqrt{|g|} e^{-2\Phi}\prt\mathcal{L}_1
\eeqa
The Lagrangians  must be invariant under the coordinate transformations and under the $B$-field   gauge transformations. Using the package "xAct" \cite{Nutma:2013zea}, one finds there are 41 couplings in the bulk Lagrangian.
As we have clarified in the Introduction section, one is free to use the restricted field redefinitions, \ie
\beqa
g_{\mu\nu}&\rightarrow &g_{\mu\nu}\nn\\
B_{\mu\nu}&\rightarrow &B_{\mu\nu}+ \alpha'\delta B^{(1)}_{\mu\nu}\nn\\
\Phi &\rightarrow &\Phi+ \alpha'\delta\Phi^{(1)}\labell{gbp}
\eeqa
where the tensors  $\delta B^{(1)}_{\mu\nu}$ and $\delta\Phi^{(1)}$ include  all possible odd- and even-parity, respectively, gauge invariant terms at two-derivative level which involve only the first derivative of the massless fields, \ie,
\beqa
  \delta B^{(1)}_{\mu\nu}&=&\alpha_1H_{\mu \nu \alpha} \nabla^{\alpha}\Phi \nn\\
 \delta\Phi^{(1)}&=&\alpha_2 H_{\alpha \beta \gamma} H^{\alpha \beta \gamma} + \alpha_3  \nabla_{\alpha}\Phi\nabla^{\alpha}\Phi\labell{fr3}
\eeqa
The coefficients $\alpha_1,\alpha_2,\alpha_3$ are arbitrary parameters. When the field variables in  $\!\!\bS_0$  are changed according to the above field redefinitions,   the following   couplings  at order $\alpha'$ are produced:
\beqa
\delta\!\!\bS_0&=&-\frac{2}{\kappa^2} \int d^{26}x\sqrt{-G}e^{-2\Phi}\Big[-\frac{1}{2}H^{\alpha\beta\gamma}\nabla_\gamma\delta B^{(1)}_{\alpha\beta}+8\nabla^\alpha\Phi\nabla_{\alpha}\delta\Phi^{(1)}
\nn\\
&&\qquad\qquad\qquad\qquad\qquad-2( R  +4 \nabla_{\alpha}\Phi \nabla^{\alpha}\Phi -\frac{1}{12} H_{\alpha \beta \gamma} H^{\alpha \beta \gamma})\delta\Phi^{(1)} \Big]\labell{var}
\eeqa
No integration by parts has been used in finding the above equation.

The couplings in the bulk action $\!\!\bS_1$ which are total derivative terms can be transferred to the boundary action $\prt\!\!\bS_1$ by using the Stokes's theorem.   Moreover, the independent couplings should not be related to each other by the Bianchi identities
\beqa
 R_{\alpha[\beta\gamma\delta]}&=&0\nn\\
 \nabla_{[\mu}R_{\alpha\beta]\gamma\delta}&=&0\labell{bian}\\
\nabla_{[\mu}H_{\alpha\beta\gamma]}&=&0\nn\\
{[}\nabla,\nabla{]}\mathcal{O}-R\mathcal{O}&=&0\nn
\eeqa
Removing the above freedoms from the most general gauge invariant couplings in the bulk,  one finds that there are 17 independent couplings. In fact, if one does not use the field redefinition, it has been shown in \cite{Metsaev:1987zx} that there are 20 independent couplings. There are three parameters in the restricted field redefinition \reef{fr3}, which reduces these couplings to 17 independent couplings. These couplings in a specific scheme are
\beqa
\mathcal{L}_1&= &  a_1 H_{\alpha }{}^{\delta \epsilon } H^{\alpha \beta \gamma }
H_{\beta \delta }{}^{\varepsilon } H_{\gamma \epsilon
\varepsilon } + a_2 H_{\alpha \beta }{}^{\delta } H^{\alpha
\beta \gamma } H_{\gamma }{}^{\epsilon \varepsilon } H_{\delta
\epsilon \varepsilon } + a_3 H_{\alpha }{}^{\gamma \delta }
H_{\beta \gamma \delta } R^{\alpha \beta } + a_4
R_{\alpha \beta } R^{\alpha \beta } \labell{L1bulk}\\&& + a_5
H_{\alpha \beta \gamma } H^{\alpha \beta \gamma } R +
a_6 R^2 + a_7 R_{\alpha \beta \gamma \delta }
R^{\alpha \beta \gamma \delta } + a_8 H_{\alpha
}{}^{\delta \epsilon } H^{\alpha \beta \gamma }
R_{\beta \gamma \delta \epsilon } + a_9 R
\nabla_{\alpha }\Phi \nabla^{\alpha }\Phi\nn\\&& + a_{10}
R^{\alpha \beta } \nabla_{\beta }\nabla_{\alpha }\Phi
+ a_{11} R_{\alpha \beta } \nabla^{\alpha }\Phi \nabla^{
\beta }\Phi + a_{12} \nabla_{\alpha }\Phi \nabla^{\alpha }\Phi
\nabla_{\beta }\Phi \nabla^{\beta }\Phi + a_{13} \nabla^{\alpha
}\Phi \nabla_{\beta }\nabla_{\alpha }\Phi \nabla^{\beta
}\Phi\nn\\&& + a_{14} \nabla_{\beta }\nabla_{\alpha }\Phi
\nabla^{\beta }\nabla^{\alpha }\Phi + a_{15} \nabla_{\alpha }H^{
\alpha \beta \gamma } \nabla_{\delta }H_{\beta \gamma
}{}^{\delta } + a_{16} H_{\alpha }{}^{\beta \gamma }
\nabla^{\alpha }\Phi \nabla_{\delta }H_{\beta \gamma
}{}^{\delta } + a_{17} \nabla_{\delta }H_{\alpha \beta \gamma }
\nabla^{\delta }H^{\alpha \beta \gamma }\nn
\eeqa
 where $a_1,\cdots, a_{17}$ are  17  parameters. The background independent assumption dictates that these parameters are independent of the geometry of the spacetime.  They may be fixed by considering the background which has a circle or a torus $T^{25}$. Then the dimensional reduction of the above couplings should have the symmetry $O(1,1)$ or $O(25,25)$, respectively. These symmetries constraint the parameters in the above action. Note that as we have anticipated in the Introduction section, the independent  couplings have no term   with three derivatives. Hence, the least action principle  does  not  constrain the bulk parameters $a_1,\cdots, a_{17}$.

The boundary of the spacetime has a unite normal vector $n^{\mu}$, hence, the boundary Lagrangian  $\prt {\cal L}_1$  should include this vector and its derivatives as well as the  other tensors. Since the field redefinition freedom has been already used in the bulk action, one is not allowed to use any field redefinition in the boundary action. Removing the total derivative terms from the most general gauge invariant boundary couplings, and using the Bianchi identities and the identities corresponding to the unit vector, it has been shown in \cite{Garousi:2021cfc} that there are 38 independent couplings in the boundary action. They are
\beqa
\prt \cL_1&=&b_{1}^{} H_{\beta \gamma \delta } H^{\beta \gamma \delta }
K^{\alpha }{}_{\alpha } + b_{2}^{} H_{\alpha }{}^{\gamma \delta
} H_{\beta \gamma \delta } K^{\alpha \beta } + b_{3}^{}
K_{\alpha }{}^{\gamma } K^{\alpha \beta } K_{\beta \gamma }
+ b_{4}^{} K^{\alpha }{}_{\alpha } K_{\beta \gamma } K^{\beta
\gamma } \nn\\&&+ b_{5}^{} K^{\alpha }{}_{\alpha } K^{\beta
}{}_{\beta } K^{\gamma }{}_{\gamma } + b_{6}^{} H_{\alpha
}{}^{\delta \epsilon } H_{\beta \delta \epsilon } K^{\gamma
}{}_{\gamma } n^{\alpha } n^{\beta } + b_{7}^{} H_{\alpha
\gamma }{}^{\epsilon } H_{\beta \delta \epsilon } K^{\gamma
\delta } n^{\alpha } n^{\beta } + b_{8}^{} K^{\alpha \beta }
R_{\alpha \beta } \nn\\&&+ b_{9}^{} K^{\gamma }{}_{\gamma }
n^{\alpha } n^{\beta } R_{\alpha \beta } + b_{10}^{}
K^{\alpha }{}_{\alpha } R + b_{11}^{} K^{\gamma \delta
} n^{\alpha } n^{\beta } R_{\alpha \gamma \beta
\delta } + b_{12}^{} H^{\beta \gamma \delta } n^{\alpha }
 \nabla_{\alpha }H_{\beta \gamma \delta }\nn\\&& + b_{13}^{} K^{\beta
\gamma } n^{\alpha } \nabla_{\alpha }K_{\beta \gamma } +
b_{14}^{} K^{\beta }{}_{\beta } n^{\alpha } \nabla_{\alpha
}K^{\gamma }{}_{\gamma } + b_{15}^{} n^{\alpha }
 \nabla_{\alpha }R + b_{16}^{} H_{\beta \gamma \delta }
H^{\beta \gamma \delta } n^{\alpha } \nabla_{\alpha }\Phi \nn\\&&+
b_{17}^{} K_{\beta \gamma } K^{\beta \gamma } n^{\alpha }
 \nabla_{\alpha }\Phi + b_{18}^{} K^{\beta }{}_{\beta }
K^{\gamma }{}_{\gamma } n^{\alpha } \nabla_{\alpha }\Phi +
b_{19}^{} H_{\beta }{}^{\delta \epsilon } H_{\gamma \delta
\epsilon } n^{\alpha } n^{\beta } n^{\gamma } \nabla_{\alpha
}\Phi \nn\\&&+ b_{20}^{} n^{\alpha } n^{\beta } n^{\gamma }
R_{\beta \gamma } \nabla_{\alpha }\Phi + b_{21}^{} n^{
\alpha } R \nabla_{\alpha }\Phi + b_{22}^{} K^{\beta
}{}_{\beta } \nabla_{\alpha }\Phi \nabla^{\alpha }\Phi +
b_{23}^{} n^{\alpha } n^{\beta } \nabla_{\alpha }\Phi
 \nabla_{\beta }K^{\gamma }{}_{\gamma }\nn\\&& + b_{24}^{} K^{\gamma
}{}_{\gamma } n^{\alpha } n^{\beta } \nabla_{\alpha }\Phi
 \nabla_{\beta }\Phi + b_{25}^{} n^{\alpha } n^{\beta }
 \nabla_{\beta } \nabla_{\alpha }K^{\gamma }{}_{\gamma } +
b_{26}^{} K^{\alpha \beta } \nabla_{\beta } \nabla_{\alpha
}\Phi\nn\\&& + b_{27}^{} K^{\gamma }{}_{\gamma } n^{\alpha }
n^{\beta } \nabla_{\beta } \nabla_{\alpha }\Phi + b_{28}^{} H_{
\alpha }{}^{\gamma \delta } H_{\beta \gamma \delta }
n^{\alpha } \nabla^{\beta }\Phi + b_{29}^{} n^{\alpha }
R_{\alpha \beta } \nabla^{\beta }\Phi + b_{30}^{}
K_{\alpha \beta } \nabla^{\alpha }\Phi \nabla^{\beta }\Phi \nn\\&&+
b_{31}^{} n^{\alpha } \nabla_{\alpha }\Phi \nabla_{\beta
}\Phi \nabla^{\beta }\Phi + b_{32}^{} n^{\alpha }
 \nabla_{\beta } \nabla_{\alpha }\Phi \nabla^{\beta }\Phi +
b_{33}^{} H_{\alpha }{}^{\delta \epsilon } n^{\alpha }
n^{\beta } n^{\gamma } \nabla_{\gamma }H_{\beta \delta
\epsilon } \nn\\&&+ b_{34}^{} n^{\alpha } n^{\beta } n^{\gamma }
 \nabla_{\alpha }\Phi \nabla_{\beta }\Phi \nabla_{\gamma
}\Phi + b_{35}^{} n^{\alpha } n^{\beta } n^{\gamma } \nabla_{
\alpha }\Phi \nabla_{\gamma } \nabla_{\beta }\Phi + b_{36}^{}
n^{\alpha } n^{\beta } n^{\gamma } \nabla_{\gamma
} \nabla_{\beta } \nabla_{\alpha }\Phi \nn\\&&+ b_{37}^{} n^{\alpha }
n^{\beta } \nabla_{\beta }K_{\alpha \gamma } \nabla^{\gamma
}\Phi + b_{38}^{} n^{\alpha } n^{\beta } n^{\gamma }
n^{\delta } \nabla_{\delta } \nabla_{\gamma }K_{\alpha \beta
}\labell{L1boundary}
\eeqa
 where $b_1,\cdots, b_{38}$ are 38 background independent parameters. In the above couplings, the first derivative of the unit vector appears in the extrinsic curvature of boundary as
\beqa
K_{\mu\nu}=\nabla_{\mu}n_{\nu}\mp n_{\mu}n^{\rho}\nabla_{\rho}n_{\nu}
\eeqa
where the minus (plus)  sign is for timelike (spacelike) boundary in which $n^\mu n_\mu=1$ ($n^\mu n_\mu=-1$). It is symmetric and satisfies $n^\mu K_{\mu\nu}=0$ and $n^\mu n^\nu\nabla_{\alpha}K_{\mu\nu}=0$ which can easily be seen by writing the unit normal vector of the boundary as
\beqa
n^{\mu}=\pm(|\nabla_\alpha f\nabla^\alpha f|)^{-1/2}\nabla^\mu f\labell{nf}
\eeqa
where plus (minus)  sign is for timelike (spacelike) boundary and   the function $f$ specifies the boundary. Note that the metric in the curvatures  in the boundary couplings \reef{L1boundary} is the bulk metric $G_{\mu\nu}$ and the metric that rises the indices is the inverse  bulk metric $G^{\mu\nu}$.

 The parameters in the boundary action \reef{L1boundary}  may  be fixed by imposing the $O(1,1)$ and $O(25,25)$ symmetries when the background has a circle and a torus $T^{25}$, respectively, and by imposing the least action principle.  Note that, unlike the bulk couplings, the above boundary couplings do have  terms with two and three derivatives, hence, in extremizing the above Lagrangian one  encounters with the variation of the second and third derivatives of massless fields on the boundary which are non-zero. As a result,  the least action principle   constrains the parameters $b_1,\cdots, b_{38}$. However, as we will see there would be no constraint on top of  the T-duality constraints.

\section{Constraint from $O(1,1)$ symmetry}

We now try to fix the parameters in the  actions \reef{L1bulk} and \reef{L1boundary}. The assumption that the effective action at the critical dimension  is background independent, means that the parameters in these actions are independent of the background. Hence, to fix them we consider a specific background which has a circle.  That is,  the open manifold has the structure $M^{(26)}=M^{(25)}\times S^{(1)}$, $\prt M^{(26)}=\prt M^{(25)}\times S^{(1)}$. The manifold $M^{(26)}$ has coordinates $x^\mu=(x^a,y)$ and  its boundary  $\prt M^{(26)}$ has coordinates $\sigma^{\tilde{\mu}}=(\sigma^\ta, y)$ where $y$ is the coordinate of the circle $S^{(1)}$. The dimensionally reduced  action then should have the $O(1,1)$ symmetry. To simplify the calculation, we consider the $Z_2$-subgroup of the $O(1,1)$-group.

 The reduction of the effective actions on the circle  $S^{(1)}$ should then be invariant under the $Z_2$-transformations \cite{Garousi:2019xlf}, \ie
 \beqa
 S_{\rm eff}(\psi)+\prt S_{\rm eff}(\psi)&=&S_{\rm eff}(\psi')+\prt S_{\rm eff}(\psi')\labell{TT}
 \eeqa
where  $S_{\rm eff}$ and  $\prt S_{\rm eff}$  are  the reductions of the bulk action $\!\!\bS_{eff}$ and boundary action $\prt\!\! \bS_{\rm eff}$, respectively. In above equation $\psi$ represents all the  massless fields in the base space which are defined in the following reductions:
 \beqa
&&G_{\mu\nu}=\left(\matrix{\bg_{ab}+e^{\varphi}g_{a }g_{b }& e^{\varphi}g_{a }&\cr e^{\varphi}g_{b }&e^{\varphi}&}\right),B_{\mu\nu}= \left(\matrix{\bb_{ab}+\frac{1}{2}b_{a }g_{b }- \frac{1}{2}b_{b }g_{a }&b_{a }\cr - b_{b }&0&}\right),\nn\\&& \Phi=\bar{\phi}+\varphi/4\,,\quad n^{\mu}=(n^a,0)\labell{reduc}
\eeqa
and $\psi'$ represents its transformation under the $Z_2$-transformations  or the T-duality transformations.

In \cite{Garousi:2021cfc}, it has been shown that  the constraint \reef{TT} can be written as two separate constraints. One for the bulk couplings and the other one for the boundary couplings. These constraints for the couplings at order $\alpha'$ are \cite{Garousi:2021cfc}
\beqa
 S_1(\psi)-S_1(\psi'_0)-\Delta S_0-\frac{2}{\kappa^2}\int d^{25}x\sqrt{-\bg}\nabla_a (A_1^a e^{-2\bphi})&=&0\nn\\
\prt S_1(\psi)-\prt S_1(\psi'_0)-\Delta\prt S_0+T_1(\psi)+\frac{2}{\kappa^2}\int d^{24}\sigma\sqrt{|\tg|}n_{a} A_1^a e^{-2\bphi}&=&0\labell{S11b}
\eeqa
where $\bg$ is the determinant of the base space metric $\bg_{ab}$ and $\tg$ is the determinant  of the induced base space metric on its boundary, \ie
\beqa
\tg_{\ta\tb}&=&\frac{\prt x^{a}}{\prt \sigma^{\ta}}\frac{\prt x^{b}}{\prt \sigma^{\tb}}\bg_{ab}\labell{gtatb}
 \eeqa
In equation \reef{S11b}, $\psi_0'$ is the transformation of the base space field $\psi$ under the Buscher rules,   $A_1^a$ is a  vector made of the  massless fields in the base space at order $\alpha'$ with arbitrary coefficients, and   $T_1(\psi)$ is the most general total derivative terms in the boundary, \ie
\beqa
T_1(\psi)=-\frac{2}{\kappa^2}\int_{\prt M}d^{D-2}\sigma \sqrt{|\tg |} n_{a}\prt_{b}(e^{-2\bphi}F_1^{ab})
\labell{bstokes1}
\eeqa
where  $ F_1^{ab} $ is an  antisymmetric  tensor constructed  from the  massless fields in the base space at order $\alpha'$ with arbitrary coefficients.
In the equation \reef{S11b}, $\Delta S_0$, $\Delta\prt S_0$ are defined in the following $\alpha'$-expansions of the reduction of the leading order actions:
\beqa
S_0(\psi'_0+\alpha'\psi'_1)-S_0(\psi'_0)&=&\alpha'\Delta S_0+\cdots\nn\\
\prt S_0(\psi'_0+\alpha'\psi'_1)-\prt S_0(\psi'_0)&=&\alpha'\Delta \prt S_0+\cdots\labell{DS0}
\eeqa
where dots represent some terms at higher orders of $\alpha'$ in which we are not interested in this paper. In the above equation, $S_0$ and $\prt S_0$ are the reduction of the leading order actions \reef{baction}. The reduction of these  actions are \cite{Kaloper:1997ux,Garousi:2021cfc}
\beqa
S_0(\psi)&=& -\frac{2}{\kappa^2}\int d^{D-1}x e^{-2\bphi}\sqrt{-\bg}\Big[\bar{R}-\nabla^a\nabla_a\vp-\frac{1}{4}\nabla_a\vp \nabla^a\vp-\frac{1}{4}(e^{\vp}V^2 +e^{-\vp}W^2)\nn\\
&&\qquad\qquad\qquad\qquad\qquad+4\nabla_a\bphi\nabla^a \bphi+2\nabla_a\bphi\nabla^a\vp-\frac{1}{12}\bH_{abc}\bH^{abc}\Big]\nn\\
\prt S_0(\psi)&=&-\frac{4}{\kappa^2}\int d^{D-2}\sigma\,e^{-2\bphi}\sqrt{| \tg|}\Big[  \bg^{ab}\bK_{ab}+\frac{1}{2} n^a\nabla_a\vp\Big]\labell{redcebs}
\eeqa
 In the first equation, $V_{ab}$ is field strength of the $U(1)$ gauge field $g_{a}$, \ie $V_{ab}=\prt_{a}g_{b}-\prt_{b}g_{a}$, and $W_{\mu\nu}$ is field strength of the $U(1)$ gauge field $b_{a}$, \ie $W_{ab}=\prt_{a}b_{\nu}-\prt_{b}b_{a}$. The    three-form $\bH$ is defined as $\bH_{abc}=\hat{H}_{abc}-\frac{3}{2}g_{[a}W_{bc]}-\frac{3}{2}b_{[a}V_{bc]}$ where the three-form  $\hat{H}$ is field strength of the two-form $\bb_{ab}  $ in \reef{reduc}.  Our notation for making  the antisymmetry  is such that \eg $g_{[a}W_{bc]}=\frac{1}{3}(g_aW_{bc}-g_{b}W_{ac}-g_cW_{ba})$. Since $\bH$ is not exterior derivative of a two-form,  it satisfies  anomalous Bianchi identity, whereas the $W,V$ satisfy the ordinary Bianchi identity, \ie
 \beqa
 \prt_{[a} \bH_{bcd]}&=&-\frac{3}{2}V_{[ab}W_{cd]}\labell{anB}\\
 \prt_{[a} W_{bc]}&=&0\nn\\
  \prt_{[a} V_{bc]}&=&0\nn
 \eeqa
   In the second equation in \reef{redcebs}, the extrinsic curvature $\bK_{ab}$  is made of the covariant derivative of the base space normal vector  $n^a$.

In \reef{DS0},  $\psi_1'$ represents the corrections to the Buscher rules at order $\alpha'$, \ie
\beqa
&&\varphi'= -\varphi+\alpha'\Delta\vp
\,\,\,,\,\,g'_{a }= b_{a }+\alpha'e^{\vp/2}\Delta g_a\,\,\,,\,\, b'_{a }= g_{a }+\alpha'e^{-\vp/2}\Delta b_a \,\,\,,\,\,\nn\\
&&\bg_{ab}'=\bg_{ab} \,\,\,,\,\,\bH_{abc}'=\bH_{abc}+\alpha'\Delta\bH_{abc} \,\,\,,\,\,  \bar{\phi}'= \bar{\phi}+\alpha'\Delta\bphi\,\,\,,\,\, n_a'=n_a\labell{T22}
\eeqa
As we have clarified in the Introduction section, the metric has no correction at order $\alpha'$, and the corrections  $\Delta \vp, \Delta b_a,\Delta g_a,\Delta \bphi$ contain all contractions of the  massless fields in the base space at order $\alpha'$ which involve only the first derivative of the massless fields.  The correction $\Delta \bH_{abc}$ is related to the corrections  $\Delta g_a$,  $\Delta b_a$ through the following relation  which is resulted from the Bianchi identity \reef{anB}:
\beqa
\Delta\bH_{abc}&=&\tilde H_{abc}-3e^{-\vp/2}W_{[ab}\Delta b_{c]}-3e^{\vp/2}\Delta g_{[a}V_{bc]}
\eeqa
where $\tilde H_{abc}$ is a $U(1)\times U(1)$ gauge invariant closed 3-form at order $\alpha'$ which is odd under parity.   It has the following terms:
\beqa
\tilde H_{abc}&=&e_1\prt_{[a}W_{b}{}^dV_{c]d}+e_2\prt_{[a}\bH_{bc]d}\nabla^d\vp\labell{tH}
\eeqa
where $e_1,e_2$ and the coefficients in the corrections   $\Delta \vp, \Delta b_a,\Delta g_a,\Delta \bphi$ are  parameters that the $Z_2$-symmetry of the effective action  should fix them. The above transformations should also  form the $Z_2$-group  \cite{Garousi:2019wgz}.

It is important to note that the torsion in the base space is $\bH_{abc}$ which is not a field strength of a two-form. Hence, in the least action principle, this field and its first derivative are known on the boundary. In the variation of the leading order bulk action \reef{redcebs}, the variation $\delta \bH_{abc}$ appears. The field redefinitions of $\bH_{abc}$ that involve  the second derivatives of the massless fields, produce the variation of the second  derivative of the massless fields. Under the use of the Stokes's theorem, it produces the variation of the first derivative of the massless fields on the boundary which is zero. Hence, the corrections \reef{tH} which involve the second derivative of the massless fields $b_a$ are consistent with our proposal for the restricted corrections to the Buscher rules which should not change the data on the boundary in the base space.

 Using the reduction \reef{redcebs}, then one can calculate $\Delta S_0$ and  $\Delta \prt S_0$ from the expansion \reef{DS0} in terms of  $\psi_1'$, \ie
\beqa
\Delta S_0&=& -\frac{2 }{\kappa^2}\int d^{25}x e^{-2\bphi}\sqrt{-\bg} \,  \Big[-4 \Big(\frac{1}{2}\bar{R} +2\prt_c\bphi\prt^c\bphi -\frac{1}{8}\prt_c\vp\prt^c\vp-\frac{1}{24}\bH^2-\frac{1}{8}e^\vp V^2\nn\\
 &&-\frac{1}{8}e^{-\vp}W^2+\frac{1}{2}\nabla_c\nabla^c\vp-\prt_c\bphi\prt^c\vp\Big)\Delta\bphi+
\frac{1}{4}\Big(  e^\vp V^2-e^{-\vp}W^2\Big)\Delta\vp\nn\\
 &&+\frac{1}{2}e^{-\vp/2}\prt_b\vp W^{ab}\Delta g_a -\frac{1}{2}e^{\vp/2}\prt_b\vp V^{ab}\Delta b_a -\frac{1}{6}\bH^{abc}\Delta\bH_{abc}\nn\\
 &&+\frac{1}{2}(\prt_a\vp+4\prt_a\bphi)\nabla^a(\Delta\vp)-\nabla_a\nabla^a(\Delta\vp)-2(\prt_a\vp-4\prt_a\bphi)\nabla^a(\Delta\bphi)
 \nn\\
&&+e^{-\vp/2}W_{ab}\nabla^b(\Delta g^a)+e^{\vp/2}V_{ab}\nabla^b(\Delta b^a)\Big] \nn\\
\Delta \prt S_0&=&-\frac{2}{\kappa^2}\int d^{24}\sigma\,e^{-2\bphi}\sqrt{| \tg|}\Big[ n^a\nabla_a(\Delta\vp)-4(\nabla_an^a-\frac{1}{2}n^a\nabla_a\vp)\Delta\bphi\Big]
\eeqa
where no integration by parts has been used.  Note that if one would like to find the equations of motion in the base space, the variation of the leading order action is the same as the above equation in which $\Delta$ should be replaced by $\delta$. In particular, the variation of the bulk action against $\delta\bH_{abc}$ is given by the last term in the third line above. As we have pointed out in the previous paragraph, unlike the other massless fields, it does not involve derivative of $\delta\bH_{abc}$.

Following the same steps as those in \cite{Garousi:2021cfc}, one finds that the $Z_2$-symmetry fixes the bulk Lagrangian \reef{L1bulk} as
\beqa
\mathcal{L}_1&= &a_{1}^{} H_{\alpha }{}^{\delta \epsilon } H^{\alpha \beta
\gamma } H_{\beta \delta }{}^{\varepsilon } H_{\gamma \epsilon
\varepsilon } + (3 a_{1}^{} + \frac{1}{64} a_{10}^{} +
\frac{1}{64} a_{11}^{}) H_{\alpha \beta }{}^{\delta }
H^{\alpha \beta \gamma } H_{\gamma }{}^{\epsilon \varepsilon }
H_{\delta \epsilon \varepsilon } -  \frac{1}{16} a_{11}^{}
H_{\alpha }{}^{\gamma \delta } H_{\beta \gamma \delta }
R^{\alpha \beta }\nn\\&& + (\frac{1}{4} a_{10}^{} +
\frac{1}{4} a_{11}^{}) R_{\alpha \beta }
R^{\alpha \beta } + \frac{1}{192} a_{11}^{} H_{\alpha
\beta \gamma } H^{\alpha \beta \gamma } R -
\frac{1}{16} a_{11}^{} R^2 + 24 a_{1}^{}
R_{\alpha \beta \gamma \delta } R^{\alpha
\beta \gamma \delta }\nn\\&& + (-36 a_{1}^{} -  \frac{1}{8} a_{10}^{}
-  \frac{1}{16} a_{11}^{}) H_{\alpha }{}^{\delta \epsilon } H^{
\alpha \beta \gamma } R_{\beta \gamma \delta \epsilon
} -  \frac{1}{4} a_{11}^{} R \nabla_{\alpha }\Phi
\nabla^{\alpha }\Phi + a_{10}^{} R^{\alpha \beta }
\nabla_{\beta }\nabla_{\alpha }\Phi\nn\\&& + a_{11}^{}
R_{\alpha \beta } \nabla^{\alpha }\Phi \nabla^{\beta
}\Phi + a_{10}^{} \nabla_{\beta }\nabla_{\alpha }\Phi
\nabla^{\beta }\nabla^{\alpha }\Phi -  \frac{1}{16} a_{10}^{}
\nabla_{\alpha }H^{\alpha \beta \gamma } \nabla_{\delta
}H_{\beta \gamma }{}^{\delta }\nn\\&& + \frac{1}{8} a_{10}^{}
H_{\alpha }{}^{\beta \gamma } \nabla^{\alpha }\Phi
\nabla_{\delta }H_{\beta \gamma }{}^{\delta } + (8 a_{1}^{} +
\frac{1}{24} a_{10}^{} + \frac{1}{48} a_{11}^{})
\nabla_{\delta }H_{\alpha \beta \gamma } \nabla^{\delta
}H^{\alpha \beta \gamma }\labell{fL1}
\eeqa
 and the boundary Lagrangian \reef{L1boundary} for  timelike boundary as
\beqa
\prt \cL_1&=& b_{1}^{} H_{\beta \gamma \delta } H^{\beta \gamma \delta }
K^{\alpha }{}_{\alpha } + \frac{1}{16} (-2 a_{10}^{} -
a_{11}^{}) H_{\alpha }{}^{\gamma \delta } H_{\beta \gamma
\delta } K^{\alpha \beta } + b_{11}^{} K_{\alpha }{}^{\gamma }
K^{\alpha \beta } K_{\beta \gamma }\nn\\&& + \frac{1}{4} (a_{11}^{}
+ 96 b_{1}^{} - 2 b_{17}^{}) K^{\alpha }{}_{\alpha } K_{\beta
\gamma } K^{\beta \gamma } + (- \frac{1}{12} a_{11}^{} - 8
b_{1}^{} -  \frac{1}{6} b_{18}^{}) K^{\alpha }{}_{\alpha }
K^{\beta }{}_{\beta } K^{\gamma }{}_{\gamma }\nn\\&& -  \frac{1}{2}
b_{19}^{} H_{\alpha }{}^{\delta \epsilon } H_{\beta \delta
\epsilon } K^{\gamma }{}_{\gamma } n^{\alpha } n^{\beta } +
(a_{10}^{} + \frac{1}{2} a_{11}^{} + 12 b_{12}^{}) K^{\alpha
\beta } R_{\alpha \beta } \nn\\&&+ \frac{1}{2} (- a_{10}^{} +
48 b_{1}^{} - 24 b_{12}^{} -  b_{17}^{} + 4 b_{19}^{}) K^{\gamma
}{}_{\gamma } n^{\alpha } n^{\beta } R_{\alpha \beta
} + (- \frac{1}{8} a_{11}^{} - 12 b_{1}^{}) K^{\alpha
}{}_{\alpha } R\nn\\&& + b_{11}^{} K^{\gamma \delta }
n^{\alpha } n^{\beta } R_{\alpha \gamma \beta \delta
} + b_{12}^{} H^{\beta \gamma \delta } n^{\alpha }
\nabla_{\alpha }H_{\beta \gamma \delta } + (\frac{1}{48}
a_{11}^{} - 2 b_{1}^{}) H_{\beta \gamma \delta } H^{\beta
\gamma \delta } n^{\alpha } \nabla_{\alpha }\Phi \nn\\&&+ b_{17}^{}
K_{\beta \gamma } K^{\beta \gamma } n^{\alpha }
\nabla_{\alpha }\Phi + b_{18}^{} K^{\beta }{}_{\beta }
K^{\gamma }{}_{\gamma } n^{\alpha } \nabla_{\alpha }\Phi +
b_{19}^{} H_{\beta }{}^{\delta \epsilon } H_{\gamma \delta
\epsilon } n^{\alpha } n^{\beta } n^{\gamma } \nabla_{\alpha
}\Phi\nn\\&& + (a_{10}^{} + \frac{1}{2} a_{11}^{} + 24 b_{12}^{} +
b_{17}^{} - 4 b_{19}^{}) n^{\alpha } n^{\beta } n^{\gamma }
R_{\beta \gamma } \nabla_{\alpha }\Phi + (-
\frac{1}{4} a_{11}^{} + 24 b_{1}^{}) n^{\alpha } R
\nabla_{\alpha }\Phi\nn\\&& - 48 b_{1}^{} K^{\beta }{}_{\beta }
\nabla_{\alpha }\Phi \nabla^{\alpha }\Phi + (96 b_{1}^{} - 2
b_{18}^{}) K^{\gamma }{}_{\gamma } n^{\alpha } n^{\beta }
\nabla_{\alpha }\Phi \nabla_{\beta }\Phi \nn\\&&+ \frac{1}{2}
\bigl(4 a_{10}^{} + a_{11}^{} + 48 (-2 b_{1}^{} +
b_{12}^{})\bigr) K^{\alpha \beta } \nabla_{\beta
}\nabla_{\alpha }\Phi \nn\\&&+ (- a_{10}^{} -  \frac{1}{2} a_{11}^{}
- 24 b_{12}^{} -  b_{17}^{} + 4 b_{19}^{}) K^{\gamma }{}_{\gamma }
n^{\alpha } n^{\beta } \nabla_{\beta }\nabla_{\alpha }\Phi +
\frac{1}{8} a_{10}^{} H_{\alpha }{}^{\gamma \delta } H_{\beta
\gamma \delta } n^{\alpha } \nabla^{\beta }\Phi\nn\\&& +
\frac{1}{2} (a_{11}^{} - 96 b_{1}^{}) n^{\alpha }
R_{\alpha \beta } \nabla^{\beta }\Phi + a_{11}^{} K_{
\alpha \beta } \nabla^{\alpha }\Phi \nabla^{\beta }\Phi + (-
a_{11}^{} + 96 b_{1}^{}) n^{\alpha } \nabla_{\alpha }\Phi
\nabla_{\beta }\Phi \nabla^{\beta }\Phi \nn\\&&+ (a_{11}^{} - 96
b_{1}^{}) n^{\alpha } \nabla_{\beta }\nabla_{\alpha }\Phi
\nabla^{\beta }\Phi + \frac{1}{8} \bigl(-2 a_{10}^{} -
a_{11}^{} - 2 (b_{11}^{} + 24 b_{12}^{})\bigr) H_{\alpha
}{}^{\delta \epsilon } n^{\alpha } n^{\beta } n^{\gamma }
\nabla_{\gamma }H_{\beta \delta \epsilon } \nn\\&&+ \frac{2}{3}
\bigl(a_{11}^{} + 2 (-96 b_{1}^{} + b_{18}^{})\bigr) n^{\alpha }
n^{\beta } n^{\gamma } \nabla_{\alpha }\Phi \nabla_{\beta
}\Phi \nabla_{\gamma }\Phi\labell{L12}\\&& + 2 (a_{10}^{} + 48 b_{1}^{} + 24
b_{12}^{} + b_{17}^{} - 4 b_{19}^{}) n^{\alpha } n^{\beta }
n^{\gamma } \nabla_{\alpha }\Phi \nabla_{\gamma
}\nabla_{\beta }\Phi + b_{38}^{} n^{\alpha } n^{\beta }
n^{\gamma } n^{\delta } \nabla_{\delta }\nabla_{\gamma
}K_{\alpha \beta }\nn
\eeqa
The bulk Lagrangian has three parameters $a_1, a_{10},a_{11}$ and the boundary Lagrangian has two bulk parameters $a_{10},a_{11}$ and 7 boundary parameters $b_1,b_{11},b_{12},b_{17},b_{18},b_{19},b_{38}$. Note that the parameter $b_{38}$ appears in only one term, hence, its corresponding couplings is invariant under the $Z_2$-transformations. Note also that the $Z_2$-symmetry does not constrain the gravity couplings in the bulk Lagrangian to be the Gauss-Bonnet combination.

 The corrections to the T-duality transformation \reef{T22} corresponding to the above Lagrangians are
\beqa
\Delta\bphi&=&- \frac{1}{128} a_{11}^{} e^{\vp} V_{ab} V^{ab} +
\frac{1}{128} a_{11} e^{-\vp}W_{ab} W^{ab} -  \frac{1}{8}
a_{11}^{} \nabla_{a}\vp \nabla^{a}\bphi \labell{dbH2}\\
  \Delta\vp&=& 24 a_{1} e^{\vp} V_ {ab} V^{ab} + 24 a_{1}  e^{-\vp}W_{ab}
W^{ab} + 48 a_{1} \nabla_{a}\vp \nabla^{a}\vp \nn\\
  \Delta g_{a}&=& 24 a_{1}e^{ \vp/2}\bH_{abc} V^{bc}+\frac{1}{4}
( a_{10}^{} +  a_{11}^{}) e^{-\vp/2}
W_{ab} \nabla^{b}\bphi + \frac{1}{16}(768 a_{1}^{} +  a_{10}^{}
+ a_{11}^{}) e^{ -\vp/2} W_{ab} \nabla^{b}\vp\nn\\
   \Delta b_{a}&=&- 24 a_{1}e^{- \vp/2}\bH_{abc} W^{bc} -\frac{1}{4}
( a_{10}^{} +  a_{11}^{}) e^{\vp/2}
V_{ab} \nabla^{b}\bphi + \frac{1}{16}(768 a_{1}^{} +  a_{10}^{}
+ a_{11}^{}) e^{ \vp/2} V_{ab} \nabla^{b}\vp\nn\\
   \Delta\bH_{abc}&=&-144 a_{1}^{}\prt_{[a}W_{b}{}^{d} V_{c]d} + (-
\frac{3}{8} a_{10}^{} -  \frac{3}{8} a_{11}^{}) \prt_{[a}\bH_{bc]d}
\nabla^{d}\vp-3e^{\vp/2} V_{[ab}\Delta g_{c]}-3e^{-\vp/2} W_{[ab}\Delta b_{c]}
\nn
\eeqa
which involve  the  three bulk parameters $a_1, a_{10},a_{11}$. In \cite{Garousi:2021cfc}, the field redefinition \reef{fr3} has not been used to write the bulk couplings in the minimal scheme, and a particular correction to the Buscher rules has been used that does not include $\Delta\bphi$, as in \cite{Kaloper:1997ux}. Note that if one sets $a_{11}$ to zero, then as we will see in the next section, the bulk couplings would not be consistent with the $O(25,25)$ symmetry whereas the couplings in \cite{Garousi:2021cfc} are consistent with the $O(25,25)$ symmetry. Hence, as expected, the corrections to the Buscher rules depend on the scheme that  one uses for the gauge invariant couplings.

\section{Constraint from $O(25,25)$ symmetry}

In the previous section we have considered  the background which has a boundary and a circle independent of it. For this background the circle reduction of the effective  actions should have the symmetry $O(1,1)$. For the simplicity of the calculations we have considered only the $Z_2$-subgroup of $O(1,1)$. The constraint from this $Z_2$-symmetry, fixes the 17 bulk parameters and 38 boundary parameters in terms of three bulk and 7 boundary parameters. One may expect if one imposes the full $O(1,1)$-symmetry, then the remaining parameters would be fixed. Or one may consider the background which has a boundary and a torus $T^2$, then the dimensional reduction of the effective actions   should have the symmetry $O(2,2)$. Imposing this symmetry, one may be able to fix all parameters in \reef{L1bulk}, \reef{L1boundary} in terms of one overall factor.  We leave this calculation for the future works, however, in this section  we consider a simpler calculation. We consider the background which depends only on one coordinate $\z$, \ie all other coordinates are assumed to be  the torus $T^{25}$. If $\z=t$  in which the boundary is spacelike, then all circles in $T^{(25)}$ are spacial coordinates, and if $\z=x$ in which the boundary is timelike,  then one of the circle is along the time direction. Then the cosmological/one-dimensional  reduction of the bulk and boundary couplings in \reef{fL1}, \reef{L12} should have the symmetry $O(25,25)$. This symmetry should constrain the parameters in   \reef{fL1}, \reef{L12}. Furthermore, in this case the lower-dimensional  boundary action should be zero. This further constrains the parameters in the boundary action.

When  fields depend only on one coordinate $\z$, using the gauge symmetries it is possible to write the metric, $B$-field  and dilaton as
 \beqa
G_{\mu\nu}=\left(\matrix{\mp n^2(\z)& 0&\cr 0&G_{ij}(\z)&}\right),\, B_{\mu\nu}= \left(\matrix{0&0\cr0&B_{ij}(\z)&}\right),\,  2\Phi=\phi+\frac{1}{2}\log|\det(G_{ij})|\labell{creduce}\eeqa
where minus (plus) sign is when $\z=t$ ($\z=x$), and the lapse function $n(\z)$ can also be fixed to $n=1$ ($n=-1$) when $\z=t$ ($\z=x$). This function at the boundary is the unit vector orthogonal to the boundary. Using the above reduction, then the  cosmological reduction of the bulk action in \reef{baction} in terms of
 the generalized metric $\cS$ which is defined as
\beqa
\cS\equiv \eta \left(\matrix{G^{-1}& -G^{-1}B&\cr BG^{-1}&G-BG^{-1}B&}\!\!\!\!\!\!\right)\labell{S}
\eeqa
where $\eta$ is the  metric of the $O(25,25)$ group, \ie
\beqa
\eta&=& \left(\matrix{0& 1&\cr 1&0&}\!\!\!\!\!\!\right),
\eeqa
becomes \cite{Veneziano:1991ek,Meissner:1991zj}
\beqa
\bS_0^c&=&-\frac{2}{\kappa^2 n}\int d\z e^{-\phi}\Bigg[-\dP^2-\frac{1}{8}\tr(\dS^2)\Bigg]\labell{S0}
\eeqa
where dot represents the $\z$-derivative. The above action  is invariant under the global $O(25,25)$ transformations because the one-dimensional dilaton is invariant and the generalized metric transforms as
\beqa
\cS&\rightarrow &\Omega^T\cS\Omega
\eeqa
where $\Omega$ belong to the  $O(25,25)$ group, \ie $\Omega^T\eta\Omega=\eta$. The cosmological/one-dimensional reduction of the boundary action in \reef{baction} becomes zero \cite{Garousi:2021cfc}, \ie $\prt\!\!\bS_0^c=0$.

Using the  reductions \reef{creduce},  one finds the following cosmological/one-dimensional reduction for  the bulk action \reef{fL1}:
\beqa
 \bS_1^c&\!\!\!\!=\!\!\!\!&-\frac{2}{\kappa^2 n}\int d\z e^{-\phi}\Bigg[\frac{1}{16} (240 a_{1}^{} + a_{10}^{} + a_{11}^{})
\dB_{i}{}^{k} \dB^{ij} \dB_{j}{}^{l} \dB_{kl} + \frac{1}{64} (192
a_{1}^{} + a_{10}^{} + a_{11}^{}) \dB_{ij} \dB^{ij} \dB_{kl}
\dB^{kl}\nn\\&& + (30 a_{1}^{} + \frac{1}{16} a_{11}^{}) \dB^{ij}
\dB^{kl} \dG_{ik} \dG_{jl} + (-36 a_{1}^{} -  \frac{1}{4}
a_{10}^{} -  \frac{1}{8} a_{11}^{}) \dB_{i}{}^{k} \dB^{ij}
\dG_{j}{}^{l} \dG_{kl}\nn\\&& + \frac{1}{16} (48 a_{1}^{} + a_{10}^{} +
a_{11}^{}) \dG_{i}{}^{k} \dG^{ij} \dG_{j}{}^{l} \dG_{kl} -
\frac{1}{16} a_{11}^{} \dG^{i}{}_{i} \dG_{j}{}^{l} \dG^{jk}
\dG_{kl}\nn\\&& + \bigl(6 a_{1}^{} + \frac{1}{256} (8 a_{10}^{} + 3
a_{11}^{})\bigr) \dB_{ij} \dB^{ij} \dG_{kl} \dG^{kl} + \bigl(3
a_{1}^{} + \frac{1}{256} (4 a_{10}^{} - 5 a_{11}^{})\bigr)
\dG_{ij} \dG^{ij} \dG_{kl} \dG^{kl}\nn\\&& + \frac{11}{256} a_{11}^{}
\dG^{i}{}_{i} \dG^{j}{}_{j} \dG_{kl} \dG^{kl} + \frac{1}{32} (2
a_{10}^{} + a_{11}^{}) \dB_{i}{}^{k} \dB^{ij} \dG_{jk}
\dG^{l}{}_{l} -  \frac{1}{256} a_{11}^{} \dB_{ij} \dB^{ij}
\dG^{k}{}_{k} \dG^{l}{}_{l}\nn\\&& -  \frac{1}{8} a_{10}^{} \dB_{i}{}^{k}
\dB^{ij} \dG_{jk} \dP + \frac{1}{8} a_{10}^{} \dG_{i}{}^{k}
\dG^{ij} \dG_{jk} \dP + \frac{1}{64} a_{11}^{} \dG^{i}{}_{i}
\dG_{jk} \dG^{jk} \dP + \frac{1}{32} a_{10}^{} \dB_{ij}
\dB^{ij} \dG^{k}{}_{k} \dP\nn\\&& + \frac{1}{64} a_{11}^{}
\dG^{i}{}_{i} \dG^{j}{}_{j} \dG^{k}{}_{k} \dP + \frac{1}{64} (4
a_{10}^{} + a_{11}^{}) \dG_{ij} \dG^{ij} \dP^2 + \frac{1}{64}
a_{11}^{} \dG^{i}{}_{i} \dG^{j}{}_{j} \dP^2 -  \frac{1}{32}
a_{10}^{} \dB^{ij} \dG^{k}{}_{k} \ddB_{ij}\nn\\&& + \frac{1}{16}
a_{10}^{} \dB^{ij} \dP \ddB_{ij} + \frac{1}{16} (384 a_{1}^{}
+ a_{10}^{} + a_{11}^{}) \ddB_{ij} \ddB^{ij} + (48 a_{1}^{} +
\frac{1}{8} a_{11}^{}) \dB^{ij} \dG_{i}{}^{k} \ddB_{jk} -
\frac{1}{8} a_{10}^{} \dG^{ij} \dP \ddG_{ij} \nn\\&&-  \frac{1}{16}
a_{11}^{} \dP^2 \ddG^{i}{}_{i} + \frac{1}{16} (384 a_{1}^{} +
a_{10}^{} + a_{11}^{}) \ddG_{ij} \ddG^{ij} + (72 a_{1}^{} +
\frac{1}{4} a_{10}^{} + \frac{3}{16} a_{11}^{}) \dB_{i}{}^{k}
\dB^{ij} \ddG_{jk}\nn\\&&- (24 a_{1}^{}+ \frac{1}{8} a_{10}^{}+
\frac{1}{8} a_{11}^{}) \dG_{i}{}^{k} \dG^{ij} \ddG_{jk} +
\frac{1}{16} a_{11}^{} \dG^{i}{}_{i} \dG^{jk} \ddG_{jk} -
\frac{1}{16} a_{11}^{} \dG^{i}{}_{i} \dP \ddG^{j}{}_{j} +
\frac{1}{64} a_{11}^{} \dB_{ij} \dB^{ij} \ddG^{k}{}_{k}\nn\\&& +
\frac{1}{32} a_{11}^{} \dG_{ij} \dG^{ij} \ddG^{k}{}_{k} -
\frac{3}{64} a_{11}^{} \dG^{i}{}_{i} \dG^{j}{}_{j} \ddG^{k}{}_{k} -
 \frac{1}{8} a_{10}^{} \dG_{ij} \dG^{ij} \ddP + \frac{1}{4}
a_{10}^{} \ddP^2\Bigg]\labell{cS3}
 \eeqa
 This action is not invariant under $O(25,25)$ transformations. Some of the non-invariant terms are total  derivative terms  which should be transferred to the boundary by using the Stokes's theorem. Moreover, the action in terms of the variables $G_{ij}, B_{ij},\Phi$ is not invariant. As we have clarified in the Introduction section, it should  be invariant in terms of some other variables which involve the first  derivatives of  $G_{ij}, B_{ij},\Phi$.

To find the total derivatives terms  in \reef{cS3}, we  add all total derivative terms at order $\alpha'$  with arbitrary coefficients to \reef{cS3}. We add the following total derivative terms:
\beqa
-\frac{2}{\kappa^2 n}\int d\z\frac{d}{d\z}(e^{-\phi}\cI_1)&=&-\frac{2}{\kappa^2 n}e^{-\phi}\cI_1\labell{intc}
\eeqa
where  $\cI_1$ is   all possible    terms at three-derivative level with even parity which are constructed from the derivatives of $\phi$, $B_{ij}$, $G_{ij}$.  Using the package   "xAct", one finds there are 18 such terms.
One can also change the field variables in the bulk action \reef{cS3} as
\begin{eqnarray}
G_{ij}&\rightarrow &G_{ij}+\alpha' \delta G^{(1)}_{ij}\nn\\
B_{ij}&\rightarrow &B_{ij}+ \alpha'\delta B^{(1)}_{ij}\nn\\
\phi &\rightarrow &\phi+ \alpha'\delta\phi^{(1)}\labell{gbpn}
\end{eqnarray}
where the matrices  $\delta G^{(1)}_{ij}$, $\delta B^{(1)}_{ij}$ and $\delta\phi^{(1)}$ are all possible  terms at 2-derivative level constructed from $\dP$, $\dB$, $\dG$.  The perturbations  $\delta G^{(1)}_{ij}$, $\delta\phi^{(1)}$ contain even-parity terms and $\delta B^{(1)}_{ij}$ contains odd-parity terms.

When the field variables   are changed according to the above  restricted field redefinitions,  then $\!\!\bS_0^c$ produces some couplings at order $\alpha'$ and higher. In this section we are interested in the resulting couplings at order $\alpha'$, \ie
\beqa
\delta\!\! \bS_0^c&=&-\frac{2}{\kappa^2 n}\int d\z e^{-\phi}\Bigg[
\delta \phi^{(1)}\left(-\frac{1}{4}\dB_{ij}\dB^{ij}-\frac{1}{4}\dG_{ij}\dG^{ij}+\dP^2\right)-2\dP\frac{d}{d\z}\delta \phi^{(1)}\labell{dS0c}\\
&&+\delta G^{(1)}_{ij}\left(-\frac{1}{2}\dB_k{}^j\dB^{ki}-\frac{1}{2}\dG_k{}^j\dG^{ki}\right)+\frac{1}{2}\dG^{ij}\frac{d}{d\z}\delta G^{(1)}_{ij}+\frac{1}{2}\dB^{ij}\frac{d}{d\z}\delta B^{(1)}_{ij}\Bigg]\nn
\eeqa
 We use the field redefinitions and the total derivative terms to remove all terms in the bulk action except the couplings which have  only the first derivative of $G_{ij},B_{ij}, \phi$.

Using the following field redefinitions:
 \beqa
\delta \phi^{(1)}&\!\!\!\!=\!\!\!\!&-6a_1 \dB_{ij} \dB^{ij} +6a_1 \dG^{i}{}_{i} \dG^{j}{}_{j}+24a_1 \dG^{i}{}_{i} \dP + 24a_1  \dP^2 \nn\\
\delta G^{(1)}_{ij}&\!\!\!\!=\!\!\!\!&-24a_1
\dB_{i}{}^{k} \dB_{jk} -24a_1 \dG_{i}{}^{k} \dG_{jk}\nn\\
\delta B^{(1)}_{ij}&\!\!\!\!=\!\!\!\!&24a_1
\dB_{j}{}^{k} \dG_{ik}-24a_1\dB_{i}{}^{k} \dG_{jk}+24a_1\dB_{ij}
\dG^{k}{}_{k} + 48a_1 \dB_{ij} \dP \labell{frd}
\eeqa
and    the following total derivative terms:
\beqa
\cI_1&=&24 a_{1}^{}  \dB_{i}{}^{k} \dB^{ij}
\dG_{jk} +12a_1\dG^{i}{}_{i} \dG_{jk} \dG^{jk}
-6a_1 \dB_{ij} \dB^{ij} \dG^{k}{}_{k}\nn\\&& - 6a_1 \dG^{i}{}_{i} \dG^{j}{}_{j} \dG^{k}{}_{k} -24a_1  \dB_{ij} \dB^{ij} \dP + 24a_1  \dG^{i}{}_{i} \dP^2 + 32a_1 \dP^3  \labell{I1}
\eeqa
one finds the cosmological action \reef{cS3} becomes invariant under the $O(25,25)$-transformations when there is the following relation between the bulk parameters $a_1,a_{10},a_{11}$:
\beqa
a_{11}&\!\!=\!\!&-384 a_1,\,\,\,a_{10}\,=\,0\labell{a11011}
\eeqa
Note that $a_{11}$ is not zero. Hence, in the minimal scheme that we have chosen for the bulk couplings \reef{L1bulk}, the correction $\Delta\bphi$ to the Buscher rules, \ie \reef{dbH2},  can not be zero.

 The $O(25,25)$-invariant form of the action is
 \beqa
 S_1^c&=&\!\!\bS_1^c+\delta\!\!\bS_0^c\,=\,-\frac{2}{\kappa^2 n}24a_1\int dt e^{-\phi}\Bigg[\frac{1}{16}\tr(\dS^4)-\frac{1}{64}(\tr(\dS^2))^2+\frac{1}{2}\tr(\dS^2)\dP^2-\frac{1}{3}\dP^4\Bigg]\labell{action2}
 \eeqa
where
\beqa
\tr(\dS^2)&=& 2\dB_{ij} \dB^{ij} + 2  \dG_{ij} \dG^{ij}\nn\\
\tr(\dS^4)&=&2 \dB_{i}{}^{k} \dB^{ij} \dB_{j}{}^{l} \dB_{kl} - 4 \dB^{ij}
\dB^{kl} \dG_{ik} \dG_{jl} + 8 \dB_{i}{}^{k} \dB^{ij} \dG_{j}{}^{l}
\dG_{kl} + 2 \dG_{i}{}^{k} \dG^{ij} \dG_{j}{}^{l} \dG_{kl}
\eeqa
The cosmological boundary action \reef{action2} is the one considered in \cite{Meissner:1996sa}. 

Since the lower-dimensional  boundary action at the leading order is zero, the field redefinitions \reef{frd} produce no term at order $\alpha'$. However, the total derivative terms \reef{I1} appear in the cosmological/one-dimensional  reduction of the boundary terms at order $\alpha'$.   Note that most of the terms in the total derivative terms are not consistent with the $O(25,25)$ symmetry. They must be cancelled with the anomalous terms in the cosmological reduction of the boundary couplings.

The one-dimensional  reduction of the timelike boundary couplings \reef{L12} is the following:
\beqa
\prt\!\!\bS_1^c&=&-\frac{2 }{\kappa^2 n}e^{-\phi}\Big[\frac{1}{4} \bigl( 96 a_1 -  (b_{11}^{} +
12 b_{12}^{})\bigr) \dB_{i}{}^{k} \dB^{ij} \dG_{jk} + \frac{1}{4}
\bigl(192 a_1 -  (b_{11}^{} + 12
b_{12}^{})\bigr) \dG_{i}{}^{k} \dG^{ij} \dG_{jk}\nn\\&& -12 a_1 \dG^{i}{}_{i} \dG_{jk} \dG^{jk} + 6 a_1 \dB_{ij} \dB^{ij} \dG^{k}{}_{k}+6 a_1 \dG^{i}{}_{i} \dG^{j}{}_{j} \dG^{k}{}_{k}\labell{ds1c}\\&& +
\frac{1}{2} (24 a_1 + 6 b_{1}^{} -
b_{19}^{}) \dB_{ij} \dB^{ij} \dP + \frac{1}{2} ( -24 a_1 + 6 b_{1}^{} -  b_{19}^{}) \dG_{ij} \dG^{ij} \dP \nn\\&&
-24 a_1 \dG^{i}{}_{i} \dP^2 + \frac{1}{6}
(-96 a_1 + 24 b_{1}^{} -  b_{18}^{}) \dP^3 + \frac{1}{4} ( -192 a_1 +  b_{11}^{} + 12 b_{12}^{}) \dB^{ij}
\ddB_{ij}\nn\\&& + \frac{1}{4} ( -192 a_1 + b_{11}^{} +
12 b_{12}^{}) \dG^{ij} \ddG_{ij} + \frac{1}{2} (192 a_1 - 24 b_{12}^{} -  b_{17}^{} + 4 b_{19}^{}) \dP
\ddP \Big]\nn
\eeqa
where we have also used the relations \reef{a11011}. The above action is not invariant under the $O(25,25)$ transformations. Note that the boundary parameter $b_{38}$ does not appear in the above action which indicates that its corresponding couplings is invariant under the $O(25,25)$ transformations. This coupling is also invariant under the $O(1,1)$ transformations. The boundary couplings in \reef{L12} which have the coefficient $b_1$ are also invariant under the $O(25,25)$ because this parameter does not appear in the above cosmological action.
If one adds to the above action the total derivative term \reef{intc} in which $\cI_1$ is given in \reef{I1}, one can choose the boundary parameters  such that the result becomes invariant. For the following relations between the parameters:
\beqa
b_{11}=96a_1-24b_1,\,\,\,\,b_{12}=8a_1
+2b_1,&&b_{19}=12b_1+\frac{1}{4}b_{17}\labell{b1219}
\eeqa
The boundary action becomes  $O(25,25)$-invariant  which involves the first derivative of the dilaton. The assumption that the boundary action must  be zero, \ie \reef{zeroS}, also implies  the following relations for the parameters:
\beqa
b_{17}=-96a_1-24b_1,&& b_{18}=96a_1
+24b_1\labell{bb}
\eeqa
That is, for the above relations one has $\prt\!\!\bS_1^c=0$.  The above relations, reduce the 7 boundary parameters in \reef{L12} to 2 parameters $b_1,b_{38}$. 
In the next section we show that the couplings satisfy the least action principle with no further constraint on the parameters, however,   the condition that the boundary couplings should include the Chern-Simons form fully constrains the parameters.


\section{Constraint from the Chern-Simons form}

For the spacetime manifolds which have boundary,  both the bulk and boundary actions should satisfy the least action principle, \ie $\delta(\!\!\bS_1+\prt\!\!\bS_1)=0$ with the appropriate boundary condition on the massless fields. For the effective action at order $\alpha'$, the massless fields and their first derivative must be known on the boundary \cite{Garousi:2021cfc}. The couplings in the bulk action \reef{fL1} contain terms which have at most two derivative on the massless fields. The variation of the bulk action then produces at most the variation of second derivative of the massless fields in the bulk. Using the Stokes's theorem, they appear on the boundary as the variation of the first derivative of the massless fields which are zero on the boundary. Hence, the least action does not constrain the couplings in \reef{fL1}.  The boundary couplings \reef{L12}, however, contain terms which have second derivative of the massless fields. In the least action, they produce the variation of the second derivative of the massless fields on the boundary which are not zero. To have no variation of the second derivative of the massless fields, the parameters in the boundary action may satisfy some constraints.

Inserting the relations \reef{a11011}, \reef{b1219} and \reef{bb} into the boundary action \reef{L12}, one finds that the variation of the resulting boundary action  against  the metric variation produces the following terms in the local frame in which the first partial derivative of metric is zero:
\beqa
&& 24(4a_1 +b_{1}^{}) \prt^{\alpha }\Phi f1^{\beta }
P^{\gamma \delta } \partial_{\alpha }\partial_{\beta
}\delta G_{\gamma \delta }
 -24(4a_1+b_1) \prt^{\alpha }\Phi
f1_{\alpha } f1^{\beta } f1^{\gamma } P^{\delta \epsilon }
\partial_{\beta }\partial_{\gamma }\delta G_{\delta \epsilon }\labell{varg}
\eeqa
where $f1^\alpha=\prt^\alpha f$. We have used the assumption that the variation of metric and its first derivative, and their tangent derivatives   are zero, \ie $\delta G_{\alpha\beta}=\prt_\mu\delta G_{\alpha\beta}=0$ and $P^{\mu\nu}\prt_\mu\prt_\gamma\delta G_{\alpha\beta}=0$.
The above variations must be zero, up to some total derivative terms on the boundary.   One finds that , up to some total derivative terms, the above variation becomes zero  with no relation between $b_1,a_1$.
We also  find that the variation of the boundary action  against  the dilaton  and the $B$-field variations produces zero result, up to some total derivative terms. Hence, the least action principle produces no constraint on the parameters on top of the constraints that are found by the T-duality.

Inserting   the relations \reef{a11011} into the bulk action \reef{fL1}, one  finds the bulk action is fixed up to one overall parameter, \ie
\beqa
\mathcal{L}_1&= &a_{1}^{}\Big[ H_{\alpha }{}^{\delta \epsilon } H^{\alpha \beta
\gamma } H_{\beta \delta }{}^{\varepsilon } H_{\gamma \epsilon
\varepsilon }-3  H_{\alpha \beta }{}^{\delta }
H^{\alpha \beta \gamma } H_{\gamma }{}^{\epsilon \varepsilon }
H_{\delta \epsilon \varepsilon } +24
H_{\alpha }{}^{\gamma \delta } H_{\beta \gamma \delta }
R^{\alpha \beta }\nn\\&& -96 R_{\alpha \beta }
R^{\alpha \beta } -2 H_{\alpha
\beta \gamma } H^{\alpha \beta \gamma } R+24 R^2 + 24
R_{\alpha \beta \gamma \delta } R^{\alpha
\beta \gamma \delta }\nn\\&& -12 H_{\alpha }{}^{\delta \epsilon } H^{
\alpha \beta \gamma } R_{\beta \gamma \delta \epsilon
} +96 R \nabla_{\alpha }\Phi
\nabla^{\alpha }\Phi  -384
R_{\alpha \beta } \nabla^{\alpha }\Phi \nabla^{\beta
}\Phi\Big] \labell{fL12}
\eeqa
The gravity part of the above couplings is the Gauss-Bonnet gravity. However, inserting the relations \reef{a11011}, \reef{b1219} and  \reef{bb} into the boundary action, one finds that the boundary action  is not fixed up to the bulk parameter $a_1$, \ie it has  two  boundary parameters $b_{1},b_{38}$ which should be fixed by some other constraints.
 One may consider other backgrounds to fix the remaining  parameters. For example, one may consider the background which has a boundary and a torus $T^2$. Then the dimensional reduction of the effective actions on this torus should have the symmetry $O(2,2)$. The bulk couplings \reef{fL12} should satisfy this symmetry automatically because there is only one overall parameter in the bulk action, however, the boundary couplings may satisfy this symmetry for some specific relations between the parameters $a_1,,b_{1}, b_{38}$. We leave this calculation for the future works.

Here we fix the remaining parameters in the boundary action by noting that the boundary couplings include the structures as those in the Chern-Simons form. Hence, we fix the remaining parameters in the boundary action  such that  the gravity couplings in the boundary  include the Chern-Simons form. The Chern-Simons form has the following gravity couplings for timelike boundary\cite{Myers:1987yn}:
\beqa
Q_2&=&4\Bigg[K^\mu{}_{\mu}\tR-2K^{\mu\nu}\tR_{\mu\nu}+\frac{1}{3}(3K^\alpha{}_{\alpha}K_{\mu\nu}K^{\mu\nu}-K^\mu{}_{\mu}K^\nu{}_{\nu}K^\alpha{}_{\alpha}-2K_{\mu}{}^{\nu}K_{\nu\alpha}K^{\alpha\mu})\Bigg]\labell{Q20}
\eeqa
where $\tR_{\mu\nu}$ and $\tR$ are curvatures that are constructed from the induced metric \reef{indg}. In terms of the spacetime metric and curvature, it is
\beqa
Q_2&=&4\Bigg[K^\mu{}_{\mu}R-2K^{\mu\nu}R_{\mu\nu}-2K_\alpha{}^\alpha n^\mu n^\nu R_{\mu\nu}+2K^{\mu\nu}n^\alpha n^\beta R_{\alpha\mu\beta\nu}\nn\\&&\qquad-\frac{1}{3}(6K^\alpha{}_{\alpha}K_{\mu\nu}K^{\mu\nu}-2K^\mu{}_{\mu}K^\nu{}_{\nu}K^\alpha{}_{\alpha}-4K_{\mu}{}^{\nu}K_{\nu\alpha}K^{\alpha\mu})\Bigg]\labell{Q2}
\eeqa
Using the following identity \cite{Garousi:2021cfc}:
\beqa
n^{\alpha } n^{\beta } n^{\gamma } n^{\delta }
\nabla_{\delta }\nabla_{\gamma }K_{\alpha \beta }&=&-2 K_{\alpha }{}^{\gamma } K^{\alpha \beta } K_{\beta
\gamma }+n^{\alpha } n^{\beta }
\nabla_{\gamma }\nabla^{\gamma }K_{\alpha \beta }\labell{iden}
\eeqa
one finds that the gravity couplings in the boundary action become the same as the couplings in $Q_2$ for the following relations:
\beqa
b_1=-4a_1,\,b_{38}=32 a_1\labell{b1b11}
\eeqa
In fact, inserting the relations \reef{a11011}, \reef{b1219}, \reef{bb} and \reef{b1b11}  into the boundary action \reef{L12}, one finds
\beqa
\prt \cL_1&=& a_1\Bigg[24 Q_2+ 32 n^{\alpha }
n^{\beta } \nabla_{\gamma }\nabla^{\gamma }K_{\alpha \beta
}-4 H_{\beta \gamma \delta } H^{\beta \gamma \delta
} K^{\alpha }{}_{\alpha } + 24 H_{\alpha }{}^{\gamma \delta }
H_{\beta \gamma \delta } K^{\alpha \beta } \nn\\&&
 + 24 H_{\alpha }{}^{\delta \epsilon } H_{\beta
\delta \epsilon } K^{\gamma }{}_{\gamma } n^{\alpha }
n^{\beta }  - 48 H_{\beta
}{}^{\delta \epsilon } H_{\gamma \delta \epsilon } n^{\alpha }
n^{\beta } n^{\gamma } \nabla_{\alpha }\Phi + 192 K^{\beta
}{}_{\beta } \nabla_{\alpha }\Phi \nabla^{\alpha }\Phi\nn\\&& - 384
K^{\gamma }{}_{\gamma } n^{\alpha } n^{\beta }
\nabla_{\alpha }\Phi \nabla_{\beta }\Phi - 384 K_{\alpha
\beta } \nabla^{\alpha }\Phi \nabla^{\beta }\Phi + 256
n^{\alpha } n^{\beta } n^{\gamma } \nabla_{\alpha }\Phi
\nabla_{\beta }\Phi \nabla_{\gamma }\Phi \Bigg]\labell{fff}
\eeqa
Hence, the boundary couplings are also fixed up to the overall bulk parameter $a_1$. 

 \section{Comparing with Meissner action}

We have used the restricted field redefinitions to find the independent bulk couplings in the minimal scheme \reef{L1bulk}. Then using the assumption that the effective action is background independent, we considered the  backgrounds in which the effective actions have symmetries $O(1,1)$ and $O(25,25)$ to fix the parameters. We have also used the assumption that the gravity couplings on the boundary should include the Chern-Simons form. Then  we have found the bulk and boundary couplings up to an overall factor. Similar calculations have been done in \cite{Garousi:2021cfc} in which no field redefinition has been used and a particular T-duality transformations has been used which produces  the bulk action to be the  Meissner action \cite{Meissner:1996sa}.  The Meissner bulk action and its corresponding timelike boundary actions are
\beqa
\bS^M_1&=&-\frac{48 a_1}{\kappa^2}\int_M d^{26} x\sqrt{-G} e^{-2\Phi}\Big[R_{GB}^2+\frac{1}{24} H_{\alpha }{}^{\delta \epsilon } H^{\alpha \beta
\gamma } H_{\beta \delta }{}^{\varepsilon } H_{\gamma \epsilon
\varepsilon } -  \frac{1}{8} H_{\alpha \beta }{}^{\delta }
H^{\alpha \beta \gamma } H_{\gamma }{}^{\epsilon \varepsilon }
H_{\delta \epsilon \varepsilon }\nn\\&&  + \frac{1}{144} H_{\alpha
\beta \gamma } H^{\alpha \beta \gamma } H_{\delta \epsilon
\varepsilon } H^{\delta \epsilon \varepsilon }+ H_{\alpha }{}^{
\gamma \delta } H_{\beta \gamma \delta } R^{\alpha
\beta } -  \frac{1}{6} H_{\alpha \beta \gamma } H^{\alpha \beta
\gamma } R  -
\frac{1}{2} H_{\alpha }{}^{\delta \epsilon } H^{\alpha \beta
\gamma } R_{\beta \gamma \delta \epsilon }\nn\\&& -
\frac{2}{3} H_{\beta \gamma \delta } H^{\beta \gamma \delta }
\nabla_{\alpha }\nabla^{\alpha }\Phi + \frac{2}{3} H_{\beta
\gamma \delta } H^{\beta \gamma \delta } \nabla_{\alpha }\Phi
\nabla^{\alpha }\Phi + 8 R \nabla_{\alpha }\Phi
\nabla^{\alpha }\Phi + 16 \nabla_{\alpha }\Phi \nabla^{\alpha
}\Phi \nabla_{\beta }\nabla^{\beta }\Phi \nn\\&&- 16
R_{\alpha \beta } \nabla^{\alpha }\Phi \nabla^{\beta
}\Phi - 16 \nabla_{\alpha }\Phi \nabla^{\alpha }\Phi \nabla_{
\beta }\Phi \nabla^{\beta }\Phi + 2 H_{\alpha }{}^{\gamma
\delta } H_{\beta \gamma \delta } \nabla^{\beta
}\nabla^{\alpha }\Phi \Big]\nn\\
\prt\!\!\bS^M_1&=&-\frac{48 a_1}{\kappa^2}\int_{\prt M} d^{25} \sigma\sqrt{-g} e^{-2\Phi}\Bigg[ Q_2+\frac{4}{3}n^{\alpha } n^{\beta }
\nabla_{\gamma }\nabla^{\gamma }K_{\alpha \beta }+
\frac{2}{3} H_{\beta \gamma \delta } H^{\beta \gamma \delta }
n^{\alpha } \nabla_{\alpha }\Phi\nn\\&& - 2 H_{\beta }{}^{\delta
\epsilon } H_{\gamma \delta \epsilon } n^{\alpha } n^{\beta }
n^{\gamma } \nabla_{\alpha }\Phi -\frac{1}{3} H_{\beta \gamma \delta } H^{\beta
\gamma \delta } K^{\alpha }{}_{\alpha } +  H_{\alpha
}{}^{\gamma \delta } H_{\beta \gamma \delta } K^{\alpha \beta
} +  H_{\alpha
}{}^{\delta \epsilon } H_{\beta \delta \epsilon } K^{\gamma
}{}_{\gamma } n^{\alpha } n^{\beta }\nn\\&&- 16 K^{\gamma
}{}_{\gamma } n^{\alpha } n^{\beta } \nabla_{\alpha }\Phi
\nabla_{\beta }\Phi  + 16 K^{\beta }{}_{\beta }
\nabla_{\alpha }\Phi \nabla^{\alpha }\Phi- 16 K_{\alpha \beta } \nabla^{\alpha
}\Phi \nabla^{\beta }\Phi \nn\\&&- 16 n^{\alpha } \nabla_{\alpha
}\Phi \nabla_{\beta }\Phi \nabla^{\beta }\Phi +
\frac{32}{3} n^{\alpha } n^{\beta } n^{\gamma }
\nabla_{\alpha }\Phi \nabla_{\beta }\Phi \nabla_{\gamma
}\Phi\Bigg]\labell{finalb}
\eeqa
where $R_{GB}^2$ is the Gauss-Bonnet bulk couplings and $Q_2$ is the Chern-Simons boundary couplings \reef{Q2}.

In this section we are going to show that the above couplings and the couplings  \reef{fL12} and \reef{fff} are identical up to a particular restricted field redefinition. In general, under the field redefinition $\Psi\rightarrow \Psi'$, the form of actions are changes. However, up to some total derivative terms, the sum of the bulk and boundary actions in terms of variable $\Psi$ should be the same as the new actions in terms of new field $\Psi'$, \ie
 \beqa
 \bS_{\rm eff}(\Psi)+\prt \!\!\bS_{\rm eff}(\Psi)&=&\bS'_{\rm eff}(\Psi')+\prt\!\! \bS'_{\rm eff}(\Psi')\labell{TT1}
 \eeqa
 Using the $\alpha'$-expansion \reef{seff} for the couplings and the $\alpha'$-expansion for the new field variable $\Psi'$, \ie
\beqa
\Psi'&=&\sum_{m=0}^\infty\alpha'^m\Psi'_m\labell{ePsi}
\eeqa
where $\Psi'_0=\Psi$, one finds the following relations  for the bulk and boundary couplings at order $\alpha'$ which are similar to the relations \reef{S11b} in the base space:
\beqa
 \bS_1(\Psi)-\bS'_1(\Psi)-\delta \!\!\bS_0-\frac{2}{\kappa^2}\int d^{26}x\sqrt{-G}\nabla_\alpha (J_1^\alpha e^{-2\Phi})&=&0\nn\\
\prt \!\!\bS_1(\Psi)-\prt\!\! \bS'_1(\Psi)-\delta\prt \!\!\bS_0+\cT_1(\Psi)+\frac{2}{\kappa^2}\int d^{25}\sigma\sqrt{|g|}n_{\alpha} J_1^\alpha e^{-2\Phi}&=&0\labell{S11b1}
\eeqa
where $\delta \!\!\bS_0$ and $\delta\prt \!\!\bS_0$ are the variations of the leading order bulk and boundary actions, respectively. $\cT_1$ is arbitrary total derivative terms on the boundary at order $\alpha'$ and $J_1^\alpha$ is arbitrary vector at order $\alpha'$ in the bulk.

The difference between the bulk couplings in the Meissner action \reef{finalb} and the bulk couplings \reef{fL12} are the following terms:
\beqa
&&a_1\Bigg[\frac{1}{6}  H_{\alpha \beta \gamma } H^{\alpha
\beta \gamma } H_{\delta \epsilon \varepsilon } H^{\delta
\epsilon \varepsilon } - 2  H_{\alpha \beta \gamma }
H^{\alpha \beta \gamma } R - 16 H_{\beta
\gamma \delta } H^{\beta \gamma \delta } \nabla_{\alpha
}\nabla^{\alpha }\Phi + 16 H_{\beta \gamma \delta }
H^{\beta \gamma \delta } \nabla_{\alpha }\Phi \nabla^{\alpha
}\Phi\nn\\&& + 96  R \nabla_{\alpha }\Phi
\nabla^{\alpha }\Phi + 384  \nabla_{\alpha }\Phi
\nabla^{\alpha }\Phi \nabla_{\beta }\nabla^{\beta }\Phi - 384
 \nabla_{\alpha }\Phi \nabla^{\alpha }\Phi
\nabla_{\beta }\Phi \nabla^{\beta }\Phi + 48
H_{\alpha }{}^{\gamma \delta } H_{\beta \gamma \delta }
\nabla^{\beta }\nabla^{\alpha }\Phi \Bigg]\nn
\eeqa
Using the variation \reef{var}, one  finds that the above terms are removable by the following field redefinitions:
\beqa
\delta\Phi^{(1)}&=&a_1\Bigg[H_{\alpha\beta\gamma}H^{\alpha\beta\gamma}-48\nabla_\alpha\Phi\nabla^\alpha\Phi\Bigg]\nn\\
\delta B_{\alpha\beta}^{(1)}&=&-96a_1H_{\alpha\beta\gamma}\nabla^\gamma\Phi\labell{PB}
\eeqa
and by  the total derivative terms with the following vector:
\beqa
J_1^\alpha&=&a_1\Bigg[16H_{\beta\gamma\delta}H^{\beta\gamma\delta}\nabla^\alpha\Phi-384\nabla^\alpha\Phi\nabla_\beta\Phi\nabla^\beta\Phi\Bigg]\labell{JJ}
\eeqa
When the field redefinition \reef{PB} is imposed on the boundary coupling \reef{baction} at the leading order, it produces the following boundary couplings order $\alpha'$:
\beqa
\delta\prt \!\!\bS_0&=&-\frac{2}{\kappa^2}\int d^{25}\sigma\sqrt{-g}e^{-2\Phi}(-4a_1)K^\mu{}_\mu\Bigg[H_{\alpha\beta\gamma}H^{\alpha\beta\gamma}-48\nabla_\alpha\Phi\nabla^\alpha\Phi\Bigg]\labell{vK}
\eeqa
One the other hand, the difference between the boundary couplings in \reef{finalb} and the boundary couplings \reef{fff} are
\beqa
a_1\Bigg[-4  H_{\beta \gamma \delta } H^{\beta \gamma \delta }
K^{\alpha }{}_{\alpha } + 16 H_{\beta \gamma \delta }
H^{\beta \gamma \delta } n^{\alpha } \nabla_{\alpha }\Phi +
192 K^{\beta }{}_{\beta } \nabla_{\alpha }\Phi
\nabla^{\alpha }\Phi - 384n^{\alpha }
\nabla_{\alpha }\Phi \nabla_{\beta }\Phi \nabla^{\beta }\Phi \Bigg]\nn
\eeqa
which are removed by the variation \reef{vK} and the vector \reef{JJ}. No total derivative term on the boundary, \ie $\cT_1(\Psi)$,  is needed to satisfy the second equation in \reef{S11b1}. Therefore, the bulk and boundary action that we have found in this paper are the same as the couplings in \reef{finalb} up to the field redefinition \reef{PB}

 \section{Discussion}

It is known   that  the classical effective actions in string theory have the gauge invariant couplings at all orders of $\alpha'$. Hence, one is free to use the higher-derivative field redefinitions. When there is no boundary, the field redefinition is to write the massless fields in terms of the derivatives of the massless fields at all orders of $\alpha'$ with arbitrary coefficients  \cite{Metsaev:1987zx}.  In this paper we have proposed that in the presence of boundary, the higher-derivative field redefinitions should be restricted to those which do not change the data on the boundary. It has been proposed in \cite{Garousi:2021cfc} that in the least action principle in the presence of the boundary, the massless fields and their derivatives up to order $n$ must be known on the boundary for the effective action at order $\alpha'^n$. The field redefinitions should not change this information on the boundary. For example, for the effective action at order $\alpha'$, the massless fields and their first derivatives are known on the boundary. The field redefinition that does not change these data, requires that  the metric  does not change, and all other massless fields  include only  the  first derivative of the massless fields. Using  this restricted field redefinition for the effective action of the bosonic string theory at order $\alpha'$, we find there are 17 independent bulk couplings, \ie \reef{L1bulk}. There are also 38 independent boundary couplings.

 We then use the assumption that the effective action of string theory at each order of $\alpha'$ and at the critical dimension is background independent, \ie the assumption that coefficients of the above 55 independent couplings are constant. We choose the background which has a boundary and a circle. The dimensional reduction of the effective action should then have the $O(1,1)$ symmetry. We constrain the 55 constants such that the couplings  become invariant under  the $Z_2$-subgroup of the $O(1,1)$-group. The $Z_2$-transformations  are the Buscher rules plus some corrections that involve only the first derivative of the massless fields in the base space. This constraint does not fully fix all the constants. We then choose another background in which all  fields depend only on one coordinate.  In this case the cosmological/one-dimensional  reduction of the effective action should have $O(25,25)$ symmetry. We  constrain the remaining parameters such that the one-dimensional  actions become invariant under the $O(25,25)$-transformations after using some one-dimensional field redefinitions which involve only the first derivative of the massless fields in the one-dimensional base space. We then impose  the constraint that the  boundary action is zero in the scheme in which the one-dimensional bulk action involves only the first derivative terms, to further fix the coefficients of the couplings.  We find that these three set of constraints fixed the 17 bulk parameters up to one overall factor. However, 2 of the boundary parameters remain unfixed. When B-field and dilaton are zero, the resulting bulk couplings are exactly the  bulk couplings in the Gauss-Bonnet gravity. Imposing the constraint that the gravity couplings in the boundary action should  be consistent with  the Chern-Simons form, the 2 boundary parameters are also fixed in terms of the overall factor of the bulk couplings. We have found the following effective actions at order $\alpha'$ for the spacetime with timelike boundary:
 \beqa
\bS_1
&=&-\frac{48a_1}{\kappa^2} \int_M d^{26}x \sqrt{-G} e^{-2\Phi}\Bigg[R^2_{\rm GB}+\frac{1}{24} H_{\alpha }{}^{\delta \epsilon } H^{\alpha \beta
\gamma } H_{\beta \delta }{}^{\varepsilon } H_{\gamma \epsilon
\varepsilon }-\frac{1}{8}  H_{\alpha \beta }{}^{\delta }
H^{\alpha \beta \gamma } H_{\gamma }{}^{\epsilon \varepsilon }
H_{\delta \epsilon \varepsilon }\nn\\&&\qquad\qquad +
R^{\alpha \beta }H_{\alpha }{}^{\gamma \delta } H_{\beta \gamma \delta }  -\frac{1}{12} R H_{\alpha
\beta \gamma } H^{\alpha \beta \gamma }  -\frac{1}{2} H_{\alpha }{}^{\delta \epsilon } H^{
\alpha \beta \gamma } R_{\beta \gamma \delta \epsilon
}\nn\\&&\qquad\qquad +4R \nabla_{\alpha }\Phi
\nabla^{\alpha }\Phi -16
 R^{\alpha \beta }\nabla_{\alpha }\Phi \nabla_{\beta
}\Phi \Bigg]\labell{ffinal}\\
\prt\!\!\bS_1&=&-\frac{48a_1}{\kappa^2}\int_{\prt M} d^{25}\sigma\sqrt{- g}  e^{-2\Phi}\Bigg[Q_2+ \frac{4}{3} n^{\alpha }
n^{\beta } \nabla_{\gamma }\nabla^{\gamma }K_{\alpha \beta
}-\frac{1}{6} H_{\beta \gamma \delta } H^{\beta \gamma \delta
} K^{\alpha }{}_{\alpha } +  H_{\alpha }{}^{\gamma \delta }
H_{\beta \gamma \delta } K^{\alpha \beta } \nn\\&&
\qquad\qquad +  H_{\alpha }{}^{\delta \epsilon } H_{\beta
\delta \epsilon } K^{\gamma }{}_{\gamma } n^{\alpha }
n^{\beta }  - 2 H_{\beta
}{}^{\delta \epsilon } H_{\gamma \delta \epsilon } n^{\alpha }
n^{\beta } n^{\gamma } \nabla_{\alpha }\Phi + 8 K^{\beta
}{}_{\beta } \nabla_{\alpha }\Phi \nabla^{\alpha }\Phi\nn\\&&\qquad\qquad - 16
K^{\gamma }{}_{\gamma } n^{\alpha } n^{\beta }
\nabla_{\alpha }\Phi \nabla_{\beta }\Phi - 16 K_{\alpha
\beta } \nabla^{\alpha }\Phi \nabla^{\beta }\Phi + \frac{32}{3}
n^{\alpha } n^{\beta } n^{\gamma } \nabla_{\alpha }\Phi
\nabla_{\beta }\Phi \nabla_{\gamma }\Phi \Bigg]\nn
\eeqa
where $R^2_{\rm GB}$ is the Gauss-Bonnet gravity couplings and $Q_2$ is the Chern-Simons couplings \reef{Q2}. We have shown that the above couplings are the same as the Meissner action and its corresponding timelike boundary couplings \reef{finalb} up to the restricted field redefinition \reef{PB}. Unlike in the Meissner action,  in the bulk action \reef{ffinal} all fields except the metric carry only the first derivative.

It has been observed in \cite{Garousi:2021cfc} that the consistency of the effective action at order $\alpha'$ with the T-duality requires not only the values of the massless fields but also the values of their first derivatives must be known on the boundary. This makes us to use the restricted field redefinition which involves only the first derivative of the massless fields. On the other hand, if one uses the field redefinitions at order $\alpha'$ which involve the second derivatives of the massless fields, then one would find 8  independent couplings in \reef{L1bulk}. We have checked it explicitly that the bulk and boundary couplings  are not consistent with the T-duality in this case. Hence, the T-duality does not allow the values of the second derivatives   of the massless fields on the boundary to be known for the effective action at order $\alpha'$.

The cosmological action \reef{cS3} in terms of the functions $\phi$, $B_{ij}$ , $G_{ij}$ which appear in the cosmological reduction of $\Phi$, $B_{\mu}$ , $G_{\mu\nu}$ have second derivatives of the functions $\phi$, $B_{ij}$ , $G_{ij}$, \ie the cosmological action \reef{cS3} is $nonlocal$. Then in the variation of the action, the first derivative of $\delta\phi$, $\delta B_{ij}$ , $\delta G_{ij}$ appear on the boundary which are zero  according to  the boundary condition proposed in  \cite{Garousi:2021cfc} for the nonlocal effective action at order $\alpha'$  that the massless fields and their first derivative should be  known on the boundary. However, after the field redefinition \reef{gbpn}, \reef{frd}, the action \reef{action2} involves only the first derivatives of the generalized metric $\cS$ and dilaton $\phi$, \ie the cosmological action \reef{action2} becomes $local$ in the new variables. Then in the least action, only variation of   these new variables appear on the boundary. Hence, in terms of the new variables, the boundary condition should be the usual boundary condition for the local actions that only the massless fields are known on the boundary. In this case, then there should be no cosmological boundary action, \ie \reef{zeroS}. In other words, the usual boundary condition for the local cosmological actions in terms of the variables $\cS,\phi$ requires the constraint \reef{zeroS}.


We have found the effective actions \reef{ffinal} for timelike boundary by requiring that the dimensional reduction of the effective action on a circle and on a torus $T^{25}$ to be invariant under the $Z_2$-transformations, and under the $O(25,25)$ transformations, respectively. Moreover,  we have imposed the constraint from   requiring  the boundary action to have the Chern-Simons form. The assumption that effective actions at the critical dimensions are background independent then requires that if one considers the background which has a boundary and a compact sub-manifold $T^2$, then the dimensional reduction of the couplings in \reef{ffinal} should have the $O(2,2)$ symmetry, up to some restricted field redefinitions. It would be interesting to perform this calculations to check if the couplings  \reef{ffinal} are consistent with the $O(2,2)$ symmetry. The dimensional reduction of the bulk actions   at orders $\alpha'^0, \alpha'$ on torus $T^2$ have been studied in \cite{Maharana:1992my,Eloy:2020dko}. It would be also interesting to find the boundary couplings for the spacelike boundary for which the Chern-Simons form is given by \reef{Q20} with different sign for the first two terms \cite{Myers:1987yn}. It would be also interesting to extend the calculations in this paper to the order $\alpha'^2$ to find the boundary couplings at this order, \ie $\prt\!\! \bS_2$ in \reef{boundT}.



\end{document}